\documentclass[a4paper]{article}

\usepackage[margin=1in]{geometry} 

\usepackage{amsmath}
\usepackage{amsthm}
\usepackage{amssymb}
\usepackage{arydshln}
\usepackage{mathtools}
\usepackage{multirow}
\usepackage{bm}
\usepackage{pgfplots}
\usepackage{tikz}

\usepackage[utf8]{inputenc}
\usepackage{hyperref}
\hypersetup{
	unicode,
	pdfauthor={Tri P. Nguyen, Ilon Joseph, Mayya Tokman},
	pdftitle={Nyström type exponential integrators for strongly magnetized charged particle dynamics},
	pdfsubject={A simple article template},
	pdfkeywords={article, template, simple},
	pdfproducer={LaTeX},
	pdfcreator={pdflatex}
}

\usepackage[sort&compress,numbers,square]{natbib}
\theoremstyle{plain}
\newtheorem{theorem}{Theorem}

\theoremstyle{definition}

\newtheorem{remark}[theorem]{Remark}

\usepackage{graphicx, color}
\graphicspath{{fig/}}

\usepackage{algorithm, algpseudocode} 
\usepackage{mathrsfs} 

\usepackage{lipsum}

\title{Nyström Type Exponential Integrators for Strongly Magnetized Charged Particle Dynamics}
\author{Tri P. Nguyen$^1$ \and Ilon Joseph$^2$ \and Mayya Tokman$^1$}

\date{
	$^1$School of Natural Sciences, University of California, Merced \\ \texttt{\{tnguyen478, mtokman\}@ucmerced.edu}\\%
	$^2$Lawrence Livermore National Laboratory \\ \texttt{joseph5@llnl.gov}\\[2ex]%
}

\begin{document}
	\maketitle
	
	\begin{abstract}
Solving for charged particle motion in electromagnetic fields (i.e. the particle pushing problem) is a computationally intensive component of particle-in-cell (PIC) methods for plasma physics simulations. This task is especially challenging when the plasma is strongly magnetized due numerical stiffness arising from the wide range of time scales between highly oscillatory gyromotion and long term macroscopic behavior. A promising approach to solve these problems is by a class of methods known as exponential integrators that can solve linear problems exactly and are A-stable. This work extends the standard exponential integration framework to derive Nyström-type exponential integrators that integrates the Newtonian equations of motion as a second-order differential equation directly. In particular, we derive second-order and third-order Nyström-type exponential integrators for strongly magnetized particle pushing problems. Numerical experiments show that the Nyström-type exponential integators exhibit significant improvement in computation speed over the standard exponential integrators.
		
		\noindent\textbf{Keywords:} Boris Algorithm, Buneman Algorithm, Charged Particle Motion, Nyström-Type Exponential Integrator, Particle Pusher
	\end{abstract}
    \vfill
	\paragraph{Acknowledgements} This work was supported in part by the Department of Energy [Contract DE-AC52 07NA27344] and the National Science Foundation [Award Numbers 1840265, 2012875]. IM Number LLNL-JRNL-2001670.

	
\newpage
\section{Introduction}
\label{sec:intro}
The problem of solving for charged particle dynamics in electromagnetic fields (i.e. the particle pushing problem) is a key component of particle-in-cell (PIC) methods in plasma simulations. In the case of strongly magnetized plasma, charged particles gyrate about magnetic field lines in highly oscillatory gyromotion on time scales significantly faster than slow-scale particle drift motion. Such multiscale temporal behavior causes strongly magnetized particle pushing problems to be numerically stiff.

The conventional approach to numerical particle pushing discretizes the Newtonian equations of motion using finite-differences and advances the dynamical state of the particle using a time stepping algorithm. The Boris \cite{Boris} and Buneman algorithms \cite{Buneman} are the most commonly used time integrators for particle pushing \cite{Qin}. These two algorithms stagger particle position and velocity by one-half time step yielding leapfrog-like, centered finite-difference schemes with second-order accuracy. A limitation of these conventional methods is that the time step size must be sufficiently small such that the electromagnetic fields are approximately constant over the time step. For problems with large field gradients, this requirement imposes a severe restriction on the time step size resulting in excessive computational expense.

Since particle pushing is the most computationally intensive part of PIC methods, there is strong research interest in the development of more efficient time integration schemes for this problem \cite{Brackbill, Cohen, Filbet1, Filbet2, Genoni, Vu}. Among recent developments in this field, two particularly interesting examples are the energy-conserving, asymptotic preserving scheme \cite{Ricketson, ChenG} and the filtered Boris algorithm \cite{Hairer1}. The first method is a Crank-Nicolson scheme modified to include an effective force approximating the grad-$B$ force acting on the guiding center (in a gyro-averaged sense) in the velocity update such that it captures the leading-order drift motion in nonuniform magnetic fields. The filtered Boris algorithm modifies the standard Boris pusher by introducing filtered functions to more accurately resolve the fast gyromotion oscillations in particle velocity due to strong magnetic fields. Different variants of the filtered Boris algorithm can be derived depending on the choice of the filter functions and where the magnetic field is evaluated. Both the modified Crank-Nicolson scheme and the filtered Boris algorithm (for the general case of arbitrary magnetic fields) are implicit methods that are more complex to implement than the standard Boris pusher. While they have the advantage of allowing for larger time step sizes for problems with nonuniform electromagnetic fields, they come at the cost of being more computationally expensive per time step than the standard Boris algorithm. To our knowledge, it has not yet been demonstrated whether these two techniques yield overall computational savings compared to Boris and similar integrators.

An alternative approach that takes into account the fact that the problem exhibits dynamical behavior on multiple scales is a class of techniques known as multi-scale methods. Examples include the two-scale formulation \cite{Crouseilles, Chartier2018, Chartier2019, Chartier2020}, the multi-revolution composition (MRC) method \cite{Chartier2018, Chartier2020}, and the micro-macro method (MM) \cite{Chartier2020}.

Recently, exponential integrators have been proposed \cite{Li2022, Wang2021, Wang2023, Wu, Nguyen} as a new technique to solve particle pushing problems. These schemes form a class of methods that solve linear problems exactly and are A-stable, thus yielding favorable computational performance in terms of accuracy and numerical stability. The integrators of \cite{Li2022} and \cite{Wang2021} are implicit energy-conserving schemes, while the integrator of \cite{Wang2023} is an explicit symmetric scheme. In \cite{Wang2023} two structure preserving methods are presented, a symplectic integrator and an energy-preserving integrator, both of which are implicit methods. In addition, the integrators of \cite{Wang2021} and \cite{Wu} are specifically designed for problems with uniform (constant) magnetic fields. While all of the exponential integrators of \cite{Li2022, Wang2021, Wang2023, Wu} compute the numerical solutions based on values of the magnetic and electric fields, the exponential integrators of \cite{Nguyen} additionally takes into account the gradients of the electromagnetic fields when computing the solution. Moreover, the integrators in \cite{Nguyen} employ an algorithm to evaluate the matrix functions (required in any exponential integration scheme) that was shown to be significantly more computationally efficient than the approach of using Krylov subspace projection methods. Numerical experiments in \cite{Nguyen} demonstrate that exponential integrators yield superior performance for linear and weakly nonlinear problems and are competitive for strongly nonlinear problems when compared to the conventional Boris and Buneman schemes. 

While it was shown in \cite{Nguyen} that exponential methods are promising as an efficient approach to numerical integration of particle pushing problems, further improvements can be made in improving the efficiency of exponential techniques for this application. In this paper, we demonstrate that even more efficient exponential methods can be derived by taking advantage of the structure of the problem. We extend the standard exponential integration framework to derive Nyström-type methods induced by partitioning the standard exponential integrators into components corresponding to particle position $\bm{x}$ and velocity $\bm{v}$. These Nyström-type exponential integrators exploit the mathematical structure of the Newtonian formulation of particle pushing problem yielding computationally efficient methods that directly integrate the equation of motion as a second-order problem.

The organization of this article is as follows. Section 2 reviews the equations describing the motion of charged particles in electromagnetic fields. Section 3 discusses the standard exponential integration framework. Section 4 presents our approach to deriving Nyström-type exponential integrators. Numerical results are presented in Section 5. Finally, we summarize our results, present conclusions, and discuss future research.

    \section{The Particle Pushing Problem}
\subsection{Equations of Motion}
The Lorentz force equation describes the dynamics of charged particle motion in an electromagnetic field. If we denote particle mass by $m$, charge by $q$, and let $\bm{B}$ and $\bm{E}$ respectively be the magnetic and electric fields, then the force acting on the particle is:

\begin{equation}\label{Lorentzforce}
m\, \frac{d\bm{v}}{dt} = q\,(\bm{E} + \bm{v} \times \bm{B}).
\end{equation}
Since velocity $\bm{v}$ is simply the time derivative of position $\bm{x}$, the Newtonian form of the particle pushing problem is equivalently expressed by the first-order system
\begin{equation}\label{NewtonEOM}
\left\{
\begin{array}{ll}
\dfrac{d\bm{x}}{dt} & = \bm{v}, \\[1.5em]
\dfrac{d\bm{v}}{dt} & = \dfrac{q}{m}(\bm{E} + \bm{v} \times \bm{B}).
\end{array}
\right.
\end{equation}
In the context of particle simulation models of plasma physics, this is known as the particle pushing problem. Accordingly, any numerical method applied to this problem is called a particle pusher.

\subsection{Particle Motion in Electromagnetic Fields}
Note that if there is only an electric field $\bm{E}$, then equation \eqref{Lorentzforce} reduces to
\[
\frac{d\bm{v}}{dt} = \frac{q}{m}\bm{E}.
\]
Hence, the particle experiences acceleration along the direction of the electric field, where a positively charged particle accelerates parallel to $\bm{E}$ while a negatively charged particle accelerates anti-parallel to $\bm{E}$.

\begin{figure}
\begin{center}
\includegraphics[scale=0.5]{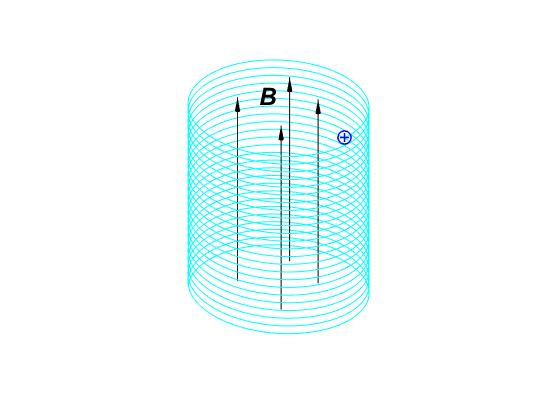}
\caption{Gyromotion in a uniform magnetic field.}\label{gyromotion}
\end{center}
\end{figure}

In the presence of a magnetic field $\bm{B}$ with no electric field, equation \eqref{Lorentzforce} simplifies to
\[
\frac{d\bm{v}}{dt} = \frac{q}{m}(\bm{v}\times\bm{B}).
\]
The cross product term implies that the magnetic field simply redirects particle velocity in a perpendicular direction without doing any net work. This results in the particle gyrating about magnetic field lines in oscillatory gyromotion in the plane perpendicular to $\bm{B}$ as illustrated in figure \ref{gyromotion}. If the magnetic field is uniform (i.e. $B = \|\bm{B}\|$ is constant in time and space), then the gyromotion has gyrofrequency $\omega = \frac{qB}{m}$, gyroperiod $T = \frac{2\pi}{\omega}$, and gyroradius $r = \frac{v_\perp}{|\omega|}$, where $v_\perp$ is particle speed perpendicular to $\bm{B}$.

If the particle is in both a magnetic field $\bm{B}$ and an electric field $\bm{E}$, then the particle experiences a type of drift motion perpendicular to both fields. As the particle goes through gyromotion from the $\bm{B}$ field, on one-half of the gyro-orbit $\bm{E}$ decelerates the particle thereby reducing its perpendicular speed $v_\perp$. Since $v_\perp$ decreases, the gyroradius $r$ decreases as well. On the other half of the gyro-orbit $\bm{E}$ accelerates the particle such that $v_\perp$ increases. This in turn increases the gyroradius $r$. The net effect on the plane perpendicular to $\bm{B}$ is that the gyromotion does not form closed orbits, but instead yields a drift motion perpendicular to both $\bm{E}_\perp$ and $\bm{B}$ as illustrated in figure \ref{ExBdriftfigure}. For the case when both the magnetic and electric fields are uniform, the $\bm{E}\times\bm{B}$ drift motion is given by \cite{Chen, Inan, Nicholson}
\begin{equation}\label{ExBdrift}
\bm{v}_E = \frac{\bm{E}\times\bm{B}}{B^2}.
\end{equation}

\begin{figure}[h]
\begin{center}
\begin{tikzpicture}
\begin{axis}[trig format plots=rad,axis equal,hide axis,xmin=-12,xmax=2.25]
\draw[thick,<-] (17.5,3)--(17.5,10) node[left] {$\bm{E}$};
\draw[thick] (17.5,10)--(17.5,17);
\draw (135,10) circle (0.1) node[below]{$\bm{B}$};
\fill (135,10) circle (1pt);
\addplot [->,domain=0:pi/2, samples=200, black] ({cos(x)-x/4}, {-sin(x)});
\addplot [->,domain=pi/2:11*pi, samples=200, black] ({cos(x)-x/4}, {-sin(x)});
\node[above,color=blue] at (130,10) {+};
\end{axis}
\end{tikzpicture}
\caption{Drift motion from an electric field perpendicular $\bm{E}$ to the $\bm{B}$ field.}\label{ExBdriftfigure}
\end{center}
\end{figure}

For the case when a particle is in a non-uniform magnetic field, the particle undergoes drift motion called grad-$B$ drift. Since the gyroradius is inversely proportional to the magnetic field strength, the gyroradius is smaller as the particle orbits in regions where the magnetic field is stronger. Conversely, the gyroradius is larger when the particle orbits in regions where the magnetic field is weaker. Similar to the $\bm{E}\times\bm{B}$ drift, the net effect is that the gyromotion does not form a closed orbit resulting in a trajectory with drift motion perpendicular to both the magnetic field $\bm{B}$ and its gradient $\nabla B$, as shown in figure \ref{gradBdrift}. Under the assumption that the change in the magnetic field is small over the gyro-orbit (i.e. $\frac{r\|\nabla B\|}{B} \ll 1$), the grad-$B$ drift is approximately given by \cite{Chen, Inan, Nicholson}
\begin{equation}
\bm{v}_\nabla = \frac{1}{2}\frac{v_\perp^2}{\omega}\frac{\bm{B}\times\nabla B}{B^2}
\end{equation}

\begin{figure}[h]
\begin{center}
\begin{tikzpicture}
\begin{axis}[trig format plots=rad,axis equal,hide axis,xmin=-12.5,xmax=3]
\draw[thick] (20,0.35)--(20,1.2) node[left] {$\nabla B$};
\draw[thick,->] (20,1.2)--(20,1.95);
\draw (145,1.2) circle (0.1) node[below]{$\bm{B}$};
\fill (145,1.2) circle (1pt);
\node[above,color=blue] at (135,1.2) {+};
\addplot[->,domain=0:pi/2,samples=100,black] ({cos(x)-x/4}, {-(1 + sin(x)/8)*sin(x)});
\addplot[->,domain=pi/2:11*pi,samples=200,black] ({cos(x)-x/4}, {-(1 + sin(x)/8)*sin(x)});
\end{axis}
\end{tikzpicture}
\caption{Grad-$B$ drift motion}\label{gradBdrift}
\end{center}
\end{figure}

The specific types of particle motion discussed above shall form the basis of test problems for the numerical experiments discussed in this work. For a discussion on particle motion due to other types of non-uniformity in the electromagnetic fields, the reader is directed to references \cite{Chen, Inan, Nicholson}.

\section{Exponential Integration}
This section reviews the framework to derive a standard exponential integrator. Consider an initial value problem of the form:
\begin{equation}\label{IVP}
\frac{d\bm{u}}{dt} = \bm{F}(\bm{u}), \qquad \bm{u}_n = \bm{u}(t_n).
\end{equation}
We take a first-order Taylor expansion of the right-hand side function about the solution $\bm{u}_n$ at time $t = t_n$ to obtain
\begin{equation}\label{1storderTaylor}
\frac{d\bm{u}}{dt} = \bm{F}(\bm{u}_n) + \bm{A}_n(\bm{u} - \bm{u}_n) + \bm{r}(\bm{u}),
\end{equation}
where $\bm{A}_n = \frac{\partial\bm{F}}{\partial\bm{u}}\big|_{\bm{u} = \bm{u}_n}$ is the Jacobian matrix (evaluated at $\bm{u} = \bm{u}_n$) and 
\begin{equation}\label{nonlinRterm}
\bm{r}(\bm{u}) = \bm{F}(\bm{u}) - \bm{F}(\bm{u}_n) - \bm{A}_n(\bm{u} - \bm{u}_n)
\end{equation}
is the nonlinear remainder term. We now multiply equation \eqref{1storderTaylor} by the integrating factor $\exp(-t\bm{A}_n)$ and integrate from $t = t_n$ to $t = t_n + h$ to obtain the variation of constants formula
\begin{equation}\label{VOC}
\bm{u}(t_n + h) = \bm{u}_n + h\,\varphi_1(h\bm{A}_n)\bm{F}(\bm{u}_n) + h\int_0^1 \exp(h(1 - \tau)\bm{A}_n)\bm{r}(\bm{u})\,d\tau,
\end{equation}
where $\varphi_1(h\bm{A}_n)$ is a matrix function defined by
\[
\varphi_1(z) = \frac{e^z - 1}{z} = \sum_{j=0}^\infty \frac{1}{(j + 1)!}z^j.
\]

Observe that if we let $\bm{u}_n$ denote a solution at time $t = t_n$ and let $h$ be a specified time step size, then equation \eqref{VOC} is an exact analytic formula for the solution at the next time step $\bm{u}(t_n + h)$. Hence, a numerical approximation to the variation of constants formula \eqref{VOC} yields an exponential integrator. For specific details on deriving exponential integrators, reference \cite{Tokman2006} discusses multistep type exponential integrators and references \cite{Tokman2006, Tokman2011} describes Runge-Kutta type exponential integrators. Note that if the problem is linear and $\bm{r(u)} = \bm{0}$, then any exponential integrator, constructed via an approximation to the nonlinear integral in \eqref{VOC},  will reduce to an exponential Euler method \cite{pope63}
\begin{equation}\label{ExpEuler}
\bm{u}_{n+1} = \bm{u}_n + h\,\varphi_1(h\bm{A}_n)\bm{F}(\bm{u}_n),
\end{equation}
where $\bm{u}_{n+1} \approx \bm{u}(t_{n+1})$ is the numerical solution at time $t_{n+1}$. If the matrix function $\varphi_1(h\bm{A}_n)$ is evaluated exactly then \eqref{ExpEuler} is simply the exact solution of the linear problem. Thus, given the definition of A-stability, any exponential method derived in this way is automatically A-stable.

\subsection{Exponential Propagation Methods, Runge-Kutta Type (EPRK)}
For the purpose of integrating the particle pushing problem, this study considers exponential integrators from a class of methods called the Exponential Propagation Iterative methods of Runge-Kutta type (EPIRK) \cite{Tokman2006}. The EPIRK class of exponential integrators has shown to be more computationally efficient per time step compared to other types of exponential integrators for numerous applications including magnetohydrodynamics (MHD) \cite{Einkemmer}.

In this study we consider two specific examples of methods from the EPIRK class. The first integrator is the second-order Exponential Propagation method (also known as an exponential Euler scheme \cite{pope63}, this method has been re-derived a number of times in the literature including \cite{Tokman2006}; e.g. see review \cite{minchevwright05})
\begin{equation}\tag{EP2}\label{EP2}
\bm{u}_{n+1} = \bm{u}_n + h\,\varphi_1(h\bm{A}_n)\bm{F}(\bm{u}_n).
\end{equation}
The second integrator is the third-order Exponential Propagation method, Runge-Kutta type \cite{Stewart, Tokman2011}
\begin{equation}\tag{EPRK3}\label{EPRK3}
\begin{array}{ccl}
\bm{U}_1 & = & \bm{u}_n + h\,\varphi_1\left(\frac{3}{4}h\bm{A}_n\right)\bm{F}(\bm{u}_n), \\[0.5em]
\bm{R}_1 & = & \bm{F}(\bm{U}_1) - \bm{F}(\bm{u}_n) - \bm{A}_n(\bm{U}_1 - \bm{u}_n), \\[0.5em]
\bm{u}_{n+1} & = & \bm{u}_n + h\,\varphi_1(h\bm{A}_n)\bm{F}(\bm{u}_n) + 2h\,\varphi_3(h\bm{A}_n)\bm{R}_1,
\end{array}
\end{equation}
where $\varphi_k$ denotes the matrix function defined by
\[
\varphi_k(z) = \displaystyle\int_0^1 \exp(z(1 - \tau))\frac{\tau^{k - 1}}{(k - 1)!} \, d\tau = \displaystyle\sum_{j=0}^\infty \frac{1}{(j + k)!}z^j, \quad k = 1, 2, \ldots .
\]
Note that if we define
\[
\varphi_0(z) \coloneq \exp(z),
\]
then for $k = 1, 2, \ldots$ the $\varphi_k$ functions are recursively defined by
\[
\varphi_k(z) = \frac{\varphi_{k-1}(z) - \varphi_{k-1}(0)}{z}.
\]
The EP2 and EPRK3 methods described above will be used in section 5 where we compare the computational performance of exponential integrators against conventional particle pushing algorithms.

\subsection{Computing the Matrix $\varphi$ Functions}
An important question is how to evaluate the matrix $\varphi$ functions required in these exponential integration schemes. The standard approach is to compute the action of the matrix function on a vector by Krylov subspace projection rather computing the matrix function. However, in \cite{Nguyen} a method to compute matrix functions called the Lagrange-Sylvester Interpolation formula \cite{Nguyen, Buchheim, Gantmacher, Sylvester} was found to be computationally efficient for low dimensional problems such as the particle pushing problem. Numerical experiments in \cite{Nguyen} show that for comparable levels of accuracy, exponential integrators using the Lagrange-Sylvester formula compute significantly faster than the same exponential integrators using Krylov subspace projection methods when applied to strongly magnetized particle pushing problems. For this reason, we choose to apply the Lagrange-Sylvester formula to compute the matrix $\varphi$ functions in this study.

The formula asserts that if $\varphi$ is a function analytic on a domain containing the spectrum of the $N \times N$ matrix $\bm{A}$, then there exists a unique $N\rm{-}1$ degree polynomial $p$ such that $p(\bm{A}) = \varphi(h\bm{A})$. More precisely, $p(\lambda)$ is the polynomial of (at most) degree $N\rm{-}1$ that interpolates $\varphi(h\lambda)$ on the eigenvalues $\lambda_1$, $\lambda_2$, $\ldots$, $\lambda_N$ of $\bm{A}$. That is, $p$ must satisfy the following linear system:
\begin{equation}\label{interpprob}
\left\{\begin{array}{ccc}
p(\lambda_1) & = & \varphi(h\lambda_1) \\[0.5em]
p(\lambda_2) & = & \varphi(h\lambda_2) \\[0.5em]
\vdots & \vdots & \vdots \\[0.5em]
p(\lambda_N) & = & \varphi(h\lambda_N)
\end{array}\right..
\end{equation}
Hence, the problem of finding the interpolation polynomial $p$ is equivalent to solving the linear system $\eqref{interpprob}$ for the unknown polynomial coefficients of $p$. If the eigenvalues $\lambda_1, \lambda_2, \ldots, \lambda_N$ are all distinct, then the interpolation problem is a system of $N$ linearly independent equations in $N$ unknowns, which is guaranteed to have a unique solution. However, if any eigenvalue is repeated, i.e. $\lambda_j$ has multiplicity $r_j > 1$, then $r_j-1$ equations are redundant for $\lambda_j$. In this case, for each repeated eigenvalue $\lambda_j$, we modify system \eqref{interpprob} by replacing the $r_j-1$ redundant equations with the following osculating conditions:
\begin{equation}\label{osculatingcond}
\begin{array}{lcll}
p'(\lambda_j) & = & \varphi'(h\lambda_j) & 1^{\rm{st}} \text{ osculating condition,} \\[0.5em]
p''(\lambda_j) & = & \varphi''(h\lambda_j) & 2^{\rm{nd}}\text{ osculating condition,} \\[0.5em]
\quad\vdots & \vdots & \quad\vdots & \quad\vdots \\[0.5em]
p^{(r_j\rm{-}1)}(\lambda_j) & = & \varphi^{(r_j\rm{-}1)}(h\lambda_j) & r_j^{\rm{th}}-1\text{ osculating condition,}
\end{array}
\end{equation}
where the superscript denotes the order of the derivative with respect to $\lambda$. This modification ensures a system of $N$ linearly independent equations in $N$ unknowns for which there is a unique solution. (For a proof of the Lagrange-Sylvester Interpolation Polynomial formula, please refer to \cite{Nguyen}.) This procedure is presented in Algorithm \ref{alg:lsf}.

\begin{algorithm} \label{alg:lsf}
\caption{Lagrange-Sylvester Formula to compute the matrix function $\varphi_k(h\bm{A})$}\label{Lagrange-Sylvester}
\begin{algorithmic}[1]
\State Solve for the eigenvalues of $\bm{A}$.
\State Solve for the interpolation polynomial $p$ such that for each eigenvalue $\lambda_j$:
\[
\begin{array}{ccc}
p(\lambda_j) & = & \varphi_k(h\lambda_j), \\[0.25em]
p'(\lambda_j) & = &\varphi_k'(h\lambda_j), \\[0.25em]
p''(\lambda_j) & = & \varphi_k''(h\lambda_j), \\[0.25em]
\vdots & \vdots & \vdots \\[0.5em]
p^{(r_j-1)}(\lambda_j) & = & \varphi_k^{(r_j-1)}(h\lambda_j),
\end{array}
\]
where $r_j \geq 1$ is the multiplicity of $\lambda_j$ and the superscript denotes the order of the derivative with respect to $\lambda$.
\State Evaluate the matrix polynomial $p(\bm{A})$.
\end{algorithmic}
\end{algorithm}

\begin{remark}\label{remark1}
It is important to note that there many possible representations for the interpolation polynomial. For example in \cite{Nguyen}, the Lagrange-Sylvestor formula calculates the interpolation polynomial of the form:
\begin{align*}
p(\lambda) & = b_0 + b_1(\lambda - \lambda_1) + b_2(\lambda - \lambda_1)(\lambda - \lambda_2) + \ldots \\
& \phantom{=} \quad + b_{N-1}(\lambda - \lambda_1)\cdots(\lambda - \lambda_{N-1}).
\end{align*}
For this particular form, known as the Newton polynomial, the polynomial coefficients are given by the Newton divided differences \cite{Burden, Kincaid}:
\[
\begin{array}{lcl}
b_0 & = & \varphi_k[\lambda_1], \\
b_1 & = & \varphi_k[\lambda_1, \lambda_2], \vspace*{0.5em}\\
b_2 & = & \varphi_k[\lambda_1, \lambda_2, \lambda_3], \vspace*{0.25em}\\
\,\,\vdots & \vdots & \qquad \vdots \vspace*{0.25em}\\
b_{N-1} & = & \varphi_k[\lambda_1, \ldots, \lambda_N].
\end{array}
\]
Here the Newton divided differences on the right-hand side are defined as follows. The zeroth divided difference is
\[
\varphi_k[\lambda_i] = \varphi_k(\lambda_i).
\]
The first divided difference is
\[
\varphi_k[\lambda_i, \lambda_{i+1}] = \left\{\begin{array}{ll}
\varphi'(\lambda_{i+1}) & \text{if } \lambda_i = \lambda_{i+1}, \vspace*{1em}\\
\dfrac{\varphi_k[\lambda_{i+1}] - \varphi_k[\lambda_i]}{\lambda_{i+1} - \lambda_i} & \text{otherwise.}
\end{array}\right.
\]
The second divided difference is
\[
\varphi_k[\lambda_i, \lambda_{i+1}, \lambda_{i+2}] = \left\{\begin{array}{ll}
\dfrac{1}{2!}\varphi_k''(\lambda_i) & \text{if }\lambda_i = \lambda_{i+1} = \lambda_{i+2}, \vspace*{1em}\\
\dfrac{\varphi_k[\lambda_{i+1, i+2}] - \varphi_k[\lambda_i, \lambda_{i+1}]}{\lambda_{i+2} - \lambda_i} & \text{otherwise.}\end{array}\right.
\]
By recursive definition, the $j$\textsuperscript{th} divided difference is
\begin{align*}
& \varphi_k[\lambda_i, \ldots, \lambda_{i+j}] \\[0.5em]
& \, = \left\{\begin{array}{ll}
\dfrac{1}{j!}\varphi_k^{(j)}(\lambda_{i+j}) & \text{if } \lambda_i, \ldots, \lambda_{i+j} \text{ are all equal,} \vspace*{1em}\\
\dfrac{\varphi_k[\lambda_{i+1}, \ldots, \lambda_{i+j}] - \varphi_k[\lambda_i, \ldots, \lambda_{i+j-1}]}{\lambda_{i+j} - \lambda_i} & \text{otherwise,}
\end{array}\right.
\end{align*}
where the superscript denotes the order of the derivative of the $\varphi_k$ function with respect to $\lambda$.
\end{remark}
 
\section{Nyström Methods}
Many dynamical systems, including the particle pushing problem under consideration here, are governed by Newton's second law of motion in which the force acting on an object is proportional to the second derivative of its position. For these systems, the governing equation of motion is expressed by a second-order initial value problem of the form:
\[
\left\{
\begin{array}{l}
\bm{x}''(t) = \bm{f}\left(\bm{x}, \bm{x}'\right), \\[0.5em]
\bm{x}(t_0) = \bm{x}_0, \qquad \bm{x}'(t_0) = \bm{x}'_0,
\end{array}
\right.
\]
where the prime notation denotes the time derivative. Conventional numerical ODE solvers typically integrate first-order initial value problems. Therefore, applying these conventional methods to solve second-order problems requires transforming them to an equivalent first-order system expressed by
\[
\frac{d}{dt}\begin{bmatrix}
\bm{x} \\
\bm{x}'
\end{bmatrix} = \begin{bmatrix}
\bm{x}' \\
\bm{f}(\bm{x},\bm{x}')
\end{bmatrix}, \qquad \begin{bmatrix}
\bm{x}(t_0) \\
\bm{x}'(t_0)
\end{bmatrix} = \begin{bmatrix}
\bm{x}_0 \\
\bm{x}'_0
\end{bmatrix},
\]
In \cite{Nystrom}  Nyström discovered a computationally efficient approach to construct integrators to solve second-order problems directly without reformulating the equations as a first order system. Such algorithms are accordingly called Nyström methods. Below we demonstrate that Nyström's idea can be used in the context of exponential integration schemes to derive more efficient exponential-Nyström methods.

\subsection{Nyström-Type Exponential Integrators}
Following Nyström's approach, we exploit the mathematical structure of the Newtonian formulation of the particle pushing problem and derive a Nyström-type exponential integrator. In particular, we employ the idea of using partitioned Runge-Kutta methods to induce Runge-Kutta-Nyström (RKN) integrators \cite{Hairer2, Leimkuhler, Sanz-Serna} and adopt this approach to derive Nyström-type exponential integrators induced by the partitioned exponential schemes.

We start by defining the function
\begin{equation}\label{fL}
    \bm{f}_L = \bm{\Omega\, v} + \frac{q}{m}\bm{E},
\end{equation}
where $\bm{\Omega}$ is the skew-symmetric matrix such that
\[
\bm{\Omega\, v} = \frac{q}{m}\bm{v}\times\bm{B}.
\]
That is,
\[
\bm{\Omega}(\bm{x}) = \dfrac{q}{m}\left\{\begin{array}{ll}
\begin{bmatrix}
\phantom{-}0 & B \\
-B & 0
\end{bmatrix} & \text{for the 2D model,} \\[1.5em]
\begin{bmatrix}
\phantom{-}0 & \phantom{-}B_z & -B_y \\
-B_z & 0 & \phantom{-}B_x \\
\phantom{-}B_y & -B_x & \phantom{-}0
\end{bmatrix} & \text{for the 3D model.}
\end{array}\right.
\]
Then equation \eqref{NewtonEOM} in the context of a particle pushing problem becomes
\[
\left\{
\begin{array}{l}
\dfrac{d\bm{x}}{dt} = \bm{v}, \\[1.5em]
\dfrac{d\bm{v}}{dt} = \bm{f}_L(\bm{x}, \bm{v}).
\end{array}
\right.
\]
Next we define the following vectors:
\[
\bm{u} = \begin{bmatrix}
\bm{x} \\
\bm{v}
\end{bmatrix}, \quad\bm{F}(\bm{u}) = \begin{bmatrix}
\bm{v} \\
\bm{f}_L(\bm{x}, \bm{v})
\end{bmatrix}, \quad\text{and}\quad\bm{u}_0 = \begin{bmatrix}
\bm{x}(t_0) \\
\bm{v}(t_0)
\end{bmatrix}.
\]
This allows us to express the particle pushing problem as an initial value problem in the form given by equation \eqref{IVP}.

We now partition the standard exponential integration framework to derive a partitioned exponential integrator scheme. Partitioning the vectors of the problem into $\bm{x}$ and $\bm{v}$ components gives us
\[
\bm{u} = \begin{bmatrix}
\bm{x} \\
\hdashline
\bm{v}
\end{bmatrix}, \qquad \bm{F}(\bm{u}) = \begin{bmatrix}
\bm{v} \\
\hdashline
\bm{f}_L(\bm{x}, \bm{v})
\end{bmatrix}, \qquad\text{and}\qquad \bm{r} = \begin{bmatrix}
\bm{r}_x \\
\hdashline
\bm{r}_v
\end{bmatrix}.
\]
Likewise, we partition the matrices of the problem into $d\times d$ block components corresponding to $\bm{x}$ and $\bm{v}$, where $d$ is the dimension of $\bm{x}$ and $\bm{v}$:
\[
\bm{A} = \left[\begin{array}{c:c}
\bm{O} & \bm{I} \\
\hdashline
\bm{H} & \bm{\Omega}
\end{array}\right], \qquad \bm{\varphi}_k(\bm{A}) = \left[\begin{array}{c:c}
\overline{\bm{\Psi}}_k(\bm{A}) & \overline{\bm{\Upsilon}}_k(\bm{A}) \\
\hdashline
\underline{\bm{\Psi}}_k(\bm{A}) & \underline{\bm{\Upsilon}}_k(\bm{A})
\end{array}\right].
\]
The block matrices expressed above are defined as follows. The blocks $\bm{O}$ and $\bm{I}$ are the zero and identity matrices, respectively. The block $\bm{H}$ is the Jacobian matrix of $\bm{f}_L$ with respect to particle position $\bm{x}$. That is,
\begin{align*}
\bm{H} & = \dfrac{\partial\bm{f}_L}{\partial\bm{x}} \\[0.5em]
& = \dfrac{\partial}{\partial\bm{x}}\left(\bm{\Omega v}\right) + \dfrac{q}{m} \dfrac{\partial\bm{E}}{\partial\bm{x}} \\[0.5em]
& = \dfrac{q}{m}\left\{\begin{array}{ll}
\begin{bmatrix}
\phantom{-}\frac{\partial B}{\partial x} v_y + \frac{\partial E_x}{\partial x} & \phantom{-}\frac{\partial B}{\partial y} v_y + \frac{\partial E_x}{\partial y} \\[1em]
-\frac{\partial B}{\partial x} v_x + \frac{\partial E_x}{\partial x} & -\frac{\partial B}{\partial y} v_x + \frac{\partial E_x}{\partial y}
\end{bmatrix} & \begin{array}{l}
\text{for the} \\
\text{2D model,}
\end{array} \\[3em]
\begin{bmatrix}
\frac{\partial B_z}{\partial x}v_y - \frac{\partial B_y}{\partial x}v_z & \frac{\partial B_z}{\partial y}v_y - \frac{\partial B_y}{\partial y}v_z & \frac{\partial B_z}{\partial z}v_y - \frac{\partial B_y}{\partial z}v_z \\[0.5em]
\frac{\partial B_x}{\partial x}v_z - \frac{\partial B_z}{\partial x}v_x & \frac{\partial B_x}{\partial y}v_z - \frac{\partial B_z}{\partial y}v_x & \frac{\partial B_x}{\partial z}v_z - \frac{\partial B_z}{\partial z}v_x \\[0.5em]
\frac{\partial B_y}{\partial x}v_x - \frac{\partial B_x}{\partial x}v_y & \frac{\partial B_y}{\partial y}v_x - \frac{\partial B_x}{\partial y}v_y & \frac{\partial B_y}{\partial z}v_x - \frac{\partial B_x}{\partial z}v_y
\end{bmatrix} \\[2.5em]
\phantom{-} + \begin{bmatrix}
\frac{\partial E_x}{\partial x} & \frac{\partial E_x}{\partial y} & \frac{\partial E_x}{\partial z} \\[0.5em]
\frac{\partial E_y}{\partial x} & \frac{\partial E_y}{\partial y} & \frac{\partial E_y}{\partial z} \\[0.5em]
\frac{\partial E_z}{\partial x} & \frac{\partial E_z}{\partial y} & \frac{\partial E_z}{\partial z}
\end{bmatrix} & \begin{array}{l}
\text{for the} \\
\text{3D model.}
\end{array}
\end{array}\right.
\end{align*}
The block matrices $\overline{\bm{\Psi}}_k$, $\overline{\bm{\Upsilon}}_k$, $\underline{\bm{\Psi}}_k$, and $\underline{\bm{\Upsilon}}_k$ respectively are the upper-left, upper-right, lower-left, and lower-right blocks of the matrix function $\varphi_k(\bm{A})$. Inserting the definition of equation \eqref{fL} into the right-hand side function $\bm{F}(\bm{u})$ yields
\[
\bm{F}(\bm{u}) = \begin{bmatrix}
\bm{v} \\
\hdashline
\bm{f}_L(\bm{x},\bm{v})
\end{bmatrix} = \begin{bmatrix}
\bm{v} \\
\hdashline
\bm{\Omega v} + \frac{q}{m}\bm{E}
\end{bmatrix}.
\]
Similarly, if we apply the definitions of the nonlinear remainder term and the block matrices of the Jacobian, then $\bm{r}$ is expressed by
\[
\bm{r} = \begin{bmatrix}
\bm{r}_x \\
\hdashline
\bm{r}_v
\end{bmatrix} = \begin{bmatrix}
\bm{0} \\
\hdashline
(\bm{\Omega} - \bm{\Omega}_n)\bm{v} + \frac{q}{m}(\bm{E} - \bm{E}_n) - \bm{H}_n(\bm{x} - \bm{x}_n)
\end{bmatrix}.
\]

Expressing the variation-of-constants formula \eqref{VOC} in vector form, we have
\begin{align*}
\begin{bmatrix}
\bm{x}(t_n + h) \\
\bm{v}(t_n + h)
\end{bmatrix} & = \begin{bmatrix}
\bm{x}_n \\
\bm{v}_n
\end{bmatrix} + h\begin{bmatrix}
\overline{\bm{\Psi}}_1(h\bm{A}_n) & \overline{\bm{\Upsilon}}_1(h\bm{A}_n) \\
\underline{\bm{\Psi}}_1(h\bm{A}_n) & \underline{\bm{\Upsilon}}_1(h\bm{A}_n)
\end{bmatrix}\begin{bmatrix}
\bm{v}_n \\
\bm{f}_L(\bm{x}_n, \bm{v}_n)
\end{bmatrix} \\
& \phantom{=} \,\, + h\int_0^1 \begin{bmatrix}
\overline{\bm{\Psi}}_0(h(1 - \tau)\bm{A}_n) & \overline{\bm{\Upsilon}}_0(h(1 - \tau)\bm{A}_n) \\
\underline{\bm{\Psi}}_0(h(1 - \tau)\bm{A}_n) & \underline{\bm{\Upsilon}}_0(h(1 - \tau)\bm{A}_n)
\end{bmatrix}\begin{bmatrix}
\bm{r}_x \\
\bm{r}_v
\end{bmatrix}\,d\tau,
\end{align*}
where
\[
\begin{bmatrix}
\overline{\bm{\Psi}}_0(h(1 - \tau)\bm{A}_n) & \overline{\bm{\Upsilon}}_0(h(1 - \tau)\bm{A}_n) \\
\underline{\bm{\Psi}}_0(h(1 - \tau)\bm{A}_n) & \underline{\bm{\Upsilon}}_0(h(1 - \tau)\bm{A}_n)
\end{bmatrix} = \varphi_0(h(1 - \tau)\bm{A}_n) \coloneq \exp(h(1 - \tau)\bm{A}_n).
\]
In other words, the analytic solutions for $\bm{x}$ and $\bm{v}$ at time $t = t_0 + h$ are:
\begin{subequations}\label{VOC_xv}
\begin{align}
\bm{x}(t_n + h) & = \bm{x}_n + h\overline{\bm{\Psi}}_1(h\bm{A}_n)\bm{v}_n + h\overline{\bm{\Upsilon}}_1(h\bm{A}_n)\bm{f}_L(\bm{x}_n, \bm{v}_n) \label{xnew}\\
& \phantom{=} \,\, + h\int_0^1 \overline{\bm{\Upsilon}}_0(h\bm{A}_n)\bm{r}_v \,d\tau, \notag\\[1em]
\bm{v}(t_n + h) & = \bm{v}_n + h\underline{\bm{\Psi}}_1(h\bm{A}_n)\bm{v}_n + h\underline{\bm{\Upsilon}}_1(h\bm{A}_n)\bm{f}_L(\bm{x}_n, \bm{v}_n) \label{vnew}\\
& \phantom{=} \,\, + h\int_0^1\underline{\bm{\Upsilon}}_n(h\bm{A}_n)\bm{r}_v \,d\tau. \notag
\end{align}
\end{subequations}

Similar to deriving a standard exponential integrator, we now let $\bm{x}_n$ and $\bm{v}_n$ denote the numerical solutions at time $t = t_n$ for position and velocity, respectively, and let $h$ be a specified time step size. Then applying appropriate quadrature rules to the nonlinear integral terms in \eqref{VOC_xv} gives us numerical approximations to the solutions $\bm{x}$ and $\bm{v}$ at the next time step. In other words, we derive Nyström-type exponential integrators induced by partitioning the standard exponential integration framework.

Since we already have formulas for second-order and third-order exponential integrators, we can readily derive schemes for second-order and third-order Runge-Kutta-Nyström-type exponential integrators. Decomposing the second-order EP2 method into $\bm{x}$ and $\bm{v}$ components gives the second-order Runge-Kutta-Nyström-type exponential integrator EPRKN2 particle pusher: 
\begin{equation}\tag{EPRKN2}\label{EPRKN2}
\begin{split}
\bm{x}_{n + 1} & = \bm{x}_n + h\,\overline{\bm{\Psi}}_1(h\bm{A}_n)\bm{v}_n + h\,\overline{\bm{\Upsilon}}_1(h\bm{A}_n)\bm{f}_L(\bm{x}_n, \bm{v}_n), \\[0.5em]
\bm{v}_{n + 1} & = \bm{v}_n + h\,\underline{\bm{\Psi}}_1(h\bm{A}_n)\bm{v}_n + h\,\underline{\bm{\Upsilon}}_1(h\bm{A}_n)\bm{f}_L(\bm{x}_n, \bm{v}_n)
\end{split}
\end{equation}
Likewise, decomposing the third-order EPRK3 exponential integrator into $\bm{x}$ and $\bm{v}$ components gives us the third-order Runge-Kutta-Nyström-type exponential integrator particle pusher:
\begin{equation}\tag{EPRKN3}\label{EPRKN3PP}
\begin{split}
\bm{X}_1 & = \bm{x}_n + h\,\overline{\bm{\Psi}}_1\left(\frac{3}{4}h\bm{A}_n\right)\bm{v}_n + h\,\overline{\bm{\Upsilon}}_1\left(\frac{3}{4}h\bm{A}_n\right)\bm{f}_L(\bm{x},\bm{v}), \\[0.5em]
\bm{V}_1 & = \bm{v}_n + h\,\underline{\bm{\Psi}}_1\left(\frac{3}{4}h\bm{A}_n\right)\bm{v}_n + h\,\underline{\bm{\Upsilon}}_1\left(\frac{3}{4}h\bm{A}_n\right)\bm{f}_L(\bm{x},\bm{v}), \\[0.5em]
\bm{R}_v & = \left(\bm{\Omega}_1 - \bm{\Omega}_n\right)\bm{V}_1 + \frac{q}{m}\big(\bm{E}(\bm{X}_1) - \bm{E}(\bm{x}_n)\big) - \bm{H}_n(\bm{X}_1 - \bm{x}_n), \\[1em]
\bm{x}_{n + 1} & = \bm{x}_n + h\,\overline{\bm{\Psi}}_1(h\bm{A}_n)\bm{v}_n + h\,\overline{\bm{\Upsilon}}_1\bm{f}_L(\bm{x}_n, \bm{v}_n) + 2h\,\overline{\bm{\Upsilon}}_3(h\bm{A}_n)\bm{R}_v, \\[0.5em]
\bm{v}_{n + 1} & = \bm{v}_n + h\,\underline{\bm{\Psi}}_1(h\bm{A}_n)\bm{v}_n + h\,\underline{\bm{\Upsilon}}_1(h\bm{A}_n)\bm{f}_L(\bm{x}_n, \bm{v}_n) + 2h\,\underline{\bm{\Upsilon}}_3(h\bm{A}_n)\bm{R}_v,
\end{split}
\end{equation}
where $\bm{\Omega}_1 = \bm{\Omega}|_{\bm{x} = \bm{X}_1}$, $\bm{\Omega}_n = \bm{\Omega}|_{\bm{x} = \bm{x}_n}$, and $\bm{H}_n = \bm{H}|_{\bm{x} = \bm{x}_n}$.

\subsection{Computing the Block Matrix Functions}
These new EPRKN2 and EPRKN3 methods will be the focus of study in our numerical experiments. Just like for the standard exponential methods \cite{Nguyen}, we still need to address the question of how the block matrix functions $\overline{\bm{\Psi}}_k, \overline{\bm{\Upsilon}}_k, \underline{\bm{\Psi}}_k, \underline{\bm{\Upsilon}}_k$ are evaluated to produce a practical integrator.

Recall that each $\varphi_k(h\bm{A})$ matrix function can be expressed by an (at most) $N-1$ degree matrix polynomial $p(\bm{A})$ by the Lagrange-Sylvester Interpolation Polynomial formula. For the specific purpose of deriving Nyström type integrators, we choose the interpolation polynomial to be of the form
\[
p(\lambda) = a_0 + a_1\lambda + a_2\lambda^2 + \ldots + a_{N-1}\lambda^{N-1},
\]
where $N = 4$ for the two dimensional model and $N = 6$ for the three dimensional model. Then the matrix function $\varphi_k(h\bm{A})$ has the polynomial expression
\[
\varphi_k(h\bm{A}) = p(\bm{A}) = a_0\bm{I}_{N\times N} + a_1\bm{A} + a_2\bm{A}^2 + \ldots + a_{N-1}\bm{A}^{N-1}.
\]
It follows that each block matrix function can then be expressed by an (at most) $N\rm{-}1$ degree matrix polynomial in terms of the blocks of $\bm{A}$. To see this, observe that the powers of $\bm{A}$ are given by the recursive formula
\[
\bm{A}^j = \begin{bmatrix}
\bm{O} & \bm{I} \\
\bm{H} & \bm{\Omega}
\end{bmatrix}^j = \begin{bmatrix}
\bm{R}^{(j-1)} & \bm{S}^{(j-1)} \\
\bm{S}^{(j-1)}\bm{H} & \bm{R}{(j-1)} + \bm{S}^{(j-1)}\bm{\Omega}
\end{bmatrix} \quad j = 1, 2, \ldots,
\]
where $\bm{R}^{(0)} = \bm{O}$ and $\bm{S}^{(0)} = \bm{I}$. Then for the two dimensional model, $\bm{A}$ is a $4 \times 4$ matrix with the matrix function $\varphi_k(h\bm{A})$ given by
\begin{align*}
\varphi_k(h\bm{A}) & = \begin{bmatrix}
\overline{\bm{\Psi}}_k(\bm{A}) & \overline{\bm{\Upsilon}}_k(\bm{A}) \\
\underline{\bm{\Psi}}_k(\bm{A}) & \underline{\bm{\Upsilon}}_k(\bm{A})
\end{bmatrix} \\[0.5em]
& = a_0\bm{I} + a_1\bm{A} + a_2\bm{A}^2 + a_3\bm{A}^3 \\
& = a_0 \begin{bmatrix}
\bm{I} & \bm{O} \\
\bm{O} & \bm{I}
\end{bmatrix} + a_1 \begin{bmatrix}
\bm{O} & \bm{I} \\
\bm{H} & \bm{\Omega}
\end{bmatrix} + a_2 \begin{bmatrix}
\bm{H} & \bm{\Omega} \\
\bm{\Omega H} & \bm{H} + \bm{\Omega}^2
\end{bmatrix} \\
& \phantom{= a_0} + a_3 \begin{bmatrix}
\bm{\Omega H} & \bm{H} + \bm{\Omega}^2 \\
(\bm{H} + \bm{\Omega}^2)\bm{\Omega} & \bm{\Omega H} + (\bm{H} + \bm{\Omega}^2)\bm{\Omega}
\end{bmatrix},
\end{align*}
where $\bm{O}, \bm{I}, \bm{H}, \bm{\Omega}$ are $2 \times 2$ matrices. This gives the following explicit expressions for the block matrix functions of the two dimensional model:
\begin{align*}
\overline{\bm{\Psi}}_k(h\bm{A}) & = a_0\bm{I} + a_2\bm{H} + a_3\bm{\Omega}\bm{H}, \\
\overline{\bm{\Upsilon}}_k(h\bm{A}) & = a_1\bm{I} + a_2\bm{\Omega} + a_3(\bm{H} + \bm{\Omega}^2), \\
\underline{\bm{\Psi}}_k(h\bm{A}) & = a_1\bm{H} + a_2\bm{\Omega H} + a_3(\bm{H} + \bm{\Omega}^2)\bm{H}, \\
\underline{\bm{\Upsilon}}_k(h\bm{A}) & = a_0\bm{I} + a_1\bm{\Omega} + a_2(\bm{H} + \bm{\Omega}^2) + a_3(\bm{\Omega H} + (\bm{H} + \bm{\Omega}^2)\bm{\Omega}).
\end{align*}
For the three dimensional model, $\bm{A}$ is a $6 \times 6$ matrix and the matrix function $\varphi_k(h\bm{A})$ is given by
\begin{align*}
\varphi_k(h\bm{A}) & = \begin{bmatrix}
\overline{\bm{\Psi}}_k(\bm{A}) & \overline{\bm{\Upsilon}}_k(\bm{A}) \\
\underline{\bm{\Psi}}_k(\bm{A}) & \underline{\bm{\Upsilon}}_k(\bm{A})
\end{bmatrix} \\[0.5em]
& = a_0\bm{I} + a_1\bm{A} + a_2\bm{A}^2 + a_3\bm{A}^3 + a_4\bm{A}^4 + a_5\bm{A}^5 \\
& = a_0 \begin{bmatrix}
\bm{I} & \bm{O} \\
\bm{O} & \bm{I}
\end{bmatrix} + a_1 \begin{bmatrix}
\bm{O} & \bm{I} \\
\bm{H} & \bm{\Omega}
\end{bmatrix} + a_2 \begin{bmatrix}
\bm{H} & \bm{\Omega} \\
\bm{R}^{(2)} & \bm{S}^{(2)}
\end{bmatrix} \\
& \phantom{=}\, + a_3 \begin{bmatrix}
\bm{R}^{(2)} & \bm{S}^{(2)} \\
\bm{R}^{(3)} & \bm{S}^{(3)}
\end{bmatrix} + a_4 \begin{bmatrix}
\bm{R}^{(3)} & \bm{S}^{(3)} \\
\bm{R}^{(4)} & \bm{S}^{(4)}
\end{bmatrix} + a_5 \begin{bmatrix}
\bm{R}^{(4)} & \bm{S}^{(4)} \\
\bm{R}^{(5)} & \bm{S}^{(5)}
\end{bmatrix},
\end{align*}
where $\bm{O}, \bm{I}, \bm{H}, \bm{\Omega}$ are $3 \times 3$ matrices and
\begin{center}
\begin{tabular}{cll}
\multirow{2}*{$\bm{R}^{(j)} = \bigg\{$} & $\bm{H}$, & $j = 1$ \\
& $\bm{S}^{(j-1)}\bm{H}$, & $j = 2,3,\ldots$ \\[0.25em]
\multirow{2}*{$\bm{S}^{(j)} = \bigg\{$} & $\bm{\Omega}$, & $j = 1$ \\
& $\bm{R}^{(j-1)} + \bm{S}^{(j-1)}\bm{\Omega}$, & $j = 2,3,\ldots$
\end{tabular}
\end{center}
This gives the polynomial expressions for the block matrix functions of the three dimensional model:
\begin{align*}
\overline{\bm{\Psi}}_k(h\bm{A}) & = a_0\bm{I} + a_2\bm{H} + a_3\bm{R}^{(2)} + a_4\bm{R}^{(3)} + a_5\bm{R}^{(4)}, \\
\overline{\bm{\Upsilon}}_k(h\bm{A}) & = a_1\bm{I} + a_2\bm{\Omega} + a_3\bm{S}^{(2)} + a_4\bm{S}^{(3)} + a_5\bm{S}^{(4)}, \\
\underline{\bm{\Psi}}_k(h\bm{A}) & = a_1\bm{H} + a_2\bm{R}^{(2)} + a_3\bm{R}^{(3)} + a_4\bm{R}^{(4)} + a_5\bm{R}^{(5)}, \\
\underline{\bm{\Upsilon}}_k(h\bm{A}) & = a_0\bm{I} + a_1\bm{\Omega} + a_2\bm{S}^{(2)} + a_3\bm{S}^{(3)} + a_4\bm{S}^{(4)} + a_5\bm{S}^{(5)}.
\end{align*}

Observe that by our choice of setting the interpolation polynomial to be of the form
\[
p(\lambda) = a_0 + a_1\lambda + a_2\lambda^2 + \ldots + a_{N-1}\lambda^{N-1},
\]
the Lagrange-Sylvester formula exploits the recursive structure of the powers of the block matrices resulting in computationally efficient polynomial expressions. Also notice that by this particular form of the polynomial, solving the interpolation problem \eqref{interpprob} is an ill-conditioned (Vandermonde) linear system. To overcome this issue, we employ Cramer's rule to derive analytic expressions for the coefficients $a_0$, $a_1$, $a_2$, $\ldots$, $a_{N-1}$ in our implementations of the Nyström exponential integrators yielding additional computational savings, see appendix \ref{coefficients}. As a final note, the analytic expressions for the coefficients $a_0$, $a_1$, $\ldots$, $a_{N\rm{-}1}$ are in general subject to catastrophic cancellation for small argument values $z = h\lambda$. Thus, for small $z = h\lambda$ we compute each coefficient using a five-term Taylor polynomial approximation for any problematic analytic expression.

\begin{remark}
We point out an important difference between the Nyström type exponential integrators discussed in this section and the standard exponential integrators when computing the exponential-like $\varphi_k$ matrix functions. Recall from remark \ref{remark1} that the matrix $\varphi_k$ functions for the standard exponential integrators of \cite{Nguyen} are computed (by the Lagrange-Sylvester Interpolation Polynomial formula) using the Newton polynomials of the form
\begin{align*}
p(\lambda) & = b_0 + b_1(\lambda - \lambda_1) + b_2(\lambda - \lambda_1)(\lambda - \lambda_2) + \ldots \\
& \phantom{=} \quad + b_{N-1}(\lambda - \lambda_1)\cdots(\lambda - \lambda_{N-1}),
\end{align*}
where the polynomial coefficients $b_0, b_1, b_2, \ldots, b_{N-1}$ are calculated by the Newton divided differences. By contrast, the Nyström type exponential integrators use the interpolation polynomial of the form
\[
p(\lambda) = a_0 + a_1\lambda + a_2\lambda^2 + \ldots + a_{N-1}\lambda^{N-1}.
\]
Here the polynomial coefficients $a_0, a_1, a_2, \ldots, a_{N-1}$ are calculated by evaluating analytic expressions of the solutions to the interpolation problem \eqref{interpprob} (along with the osculating conditions \eqref{osculatingcond} for the case when there are repeated eigenvalues in the spectrum of $\bm{A}$).
\end{remark}

\section{Numerical Experiments}
In this section, we compare the computational performances of the Runge-Kutta-Nyström-type exponential integrators EPRKN2 and EPRKN3 against the standard EP2 and EPRK3 exponential integrators. The Boris and Buneman algorithms are also included to evaluate how well these exponential integrators perform against conventional particle pushers.

All numerical experiments model a particle of unit mass and unit charge in a strong magnetic field orientated in the $z$ direction. The specific configurations for each test problem are described in the next two subsections below. For reference, we computed highly accurate solutions to each test problem using the MATLAB \texttt{ode113} integrator with error tolerances set to $10^{-12}$ for \texttt{RelTol} (relative error tolerance) and $10^{-12}$ for \texttt{AbsTol} (absolute error tolerance). The error of the numerical solution with respect to particle position is defined by
\[
\text{error, position} = \frac{\|\bm{x}^* - \bm{x}\|}{\|\bm{x}^*\|},
\]
where $\bm{x}^*$ is the particle position of the reference solution, $\bm{x}$ is the particle position of the particle pusher, and $\|\cdot\|$ denotes the Euclidean norm. Similarly, the error of the numerical solution with respect to particle velocity is defined by
\[
\text{error, velocity} = \frac{\|\bm{v}^* - \bm{v}\|}{\|\bm{v}^*\|},
\]
where $\bm{v}^*$ is the particle position of the reference solution and $\bm{v}$ is the particle position of the particle pusher. Experiments with the electric potential well problems were implemented in C++ using the Eigen C++ template library for linear algebra \cite{Eigen}. Experiments with the gyroradius and grad-$B$ drift problems were implemented in MATLAB. All computations in these experiments were calculated with double precision floating point operations.

\subsection{Two Dimensional Model Configurations}
The initial conditions for all two dimensional model test problems are $\bm{x}_0 = (1, 0)$ and $\bm{v}_0 = (0, -1)$. Each test problem is integrated over the time interval [0, 100].
\begin{itemize}
    \item \textbf{Electric Potential Well Problems:} These test problems consider particle motion in a strong uniform (constant in time and space) magnetic field $\bm{B} = 100\,\hat{\bm{z}}$ and a non-uniform (in space) electric field $\bm{E}$. The electric field $\bm{E}$ is given by an electric potential well such that the resulting anisotropic drift motion forms a closed orbit on temporal and spatial scales much larger than the gyromotion. The experiments are conducted with three specific electric fields given by a quadratic potential, a cubic potential, and a quartic potential. The electric field for each electric potential well test problem is specified in table \ref{EM2D}. This experiment is first conducted with magnetic field $\bm{B} = 100\,\hat{\bm{z}}$. In the interest of examining the influence of the magnitude of the magnetic field on the test problems, the experiment is then repeated with a stronger magnetic field  $\bm{B} = 1000\,\hat{\bm{z}}$.
    \item \textbf{Gyroradius Problem:} This experiment examines the gyroradius of the solutions computed by the numerical particle pushers. A known issue with the Boris pusher \cite{Parker} (as well as with many other conventional particle pushers such as the Buneman pusher) is that in problems with an $\bm{E} \times \bm{B}$ drift motion it computes an artificially enlarged gyroradius when using large step sizes relative to the gyroperiod. In this context, a step size $h$ is considered "small" when $\omega h \ll 1$ and "large" when $\omega h \gg 1$, where $\omega = qB/m$ is the gyroperiod. For strongly magnetized problems, this can impose a severe restriction on the step size for many conventional particle pushers if accuracy requirements of the simulations demand resolution at the scale of the gyroradius. We include this experiment to study how well the exponential integrators are able to correctly resolve the gyroradius. The test problem under consideration is a linear $\bm{E}\times\bm{B}$ drift problem with electromagnetic fields
    \[
    \bm{B} = 100\,\hat{\bm{z}} \quad\text{and}\quad \bm{E} = -\begin{bmatrix}
    0 \\
    1 + y
    \end{bmatrix},
    \]
    which has a gyroradius of $r = 0.01$. The problem is integrated using a "small" step size $h = 0.001$ and a "large" step size $h = 0.1$.
    \item \textbf{Grad-$B$ Drift Problem:} This experiment examines the so-called grad-$B$ drift problem. The test problem has a non-uniform magnetic field with a gradient term in which the length scale of the spatial variation is much longer than the gyroradius. In other words, the variation in the magnetic field that the particle experiences is "small" over the gyro-orbit. This is formally stated by
    \[
    \frac{r\|\nabla B\|}{B} \ll 1,
    \]
    where $r$ is the gyroradius, $B = \|\bm{B}\|$, and $\nabla B$ is the magnetic field gradient. Under this condition, the particle experiences an approximate drift velocity of \cite{Chen, Inan, Nicholson}
    \[
    \bm{v}_{\nabla B} = \frac{1}{2}\frac{v_\perp^2}{\omega}\frac{\bm{B}\times\nabla B}{B^2},
    \]
    where $v_\perp$ is the particle speed in the plane perpendicular to the magnetic field and $\omega = qB/m$ is the gyrofrequency. The electromagnetic fields for this test problem are
    \[
    \bm{B} = (100 + \delta B\,y)\hat{\bm{z}} \quad\text{and}\quad \bm{E} = \bm{0}.
    \]
\end{itemize}

\begin{table}[H]
\centering
\begin{tabular}{lccc}
& Quadratic Well & Cubic Well & Quartic Well \\[1em]
Potential $V(\bm{x})$: & $50(x^2 + y^2)$ & $47(x^2 + y^2)$ & $\frac{25}{3}(x^4 + y^4)$ \\
& & $+ x^3 + y^3$ \\[1em]
Electric Field $\bm{E}(\bm{x})$: & $-100\begin{bmatrix}
x \\ y
\end{bmatrix}$ & $-\begin{bmatrix}
94x + 3x^2 \\ 94y + 3y^2
\end{bmatrix}$ & $-\frac{100}{3}\begin{bmatrix}
x^3 \\ y^3
\end{bmatrix}$
\end{tabular}
\caption{Electric fields for 2D potential well problems with uniform magnetic field $\bm{B} = 100\,\hat{\bm{z}}$}\label{EM2D}
\end{table}

\subsection{Three Dimensional Model Configurations}
All three dimensional test problems are electric potential well configurations with a uniform magnetic field $\bm{B} = 100\,\hat{\bm{z}}$ and a spatially non-uniform electric field $\bm{E}$. Similar to the two dimensional model, we examine three specific electric fields given by a quadratic potential, a cubic potential, and a quartic potential. Configurations for the electric scalar potential wells and their corresponding electric fields are shown in table \ref{EM3D}. The initial conditions are $\bm{x}_0 = (1, 0, 0)$ and $\bm{v}_0 = (0, -1, 1)$. Each test problem is integrated over the time interval [0, 100].

\begin{table}[H]
\centering
\begin{tabular}{lccc}
& Quadratic Well & Cubic Well & Quartic Well \\[1em]
\multirow{3}{*}{Potential $V(\bm{x})$:} & $50(x^2 + y^2)$ & $47(x^2 + y^2)$ & $\frac{25}{3}(x^4 + y^4)$ \\
& $+ 5z^2$ & $+ x^3 + y^3$ & $+ \frac{5}{6}z^4$ \\
& & $ + \frac{1}{10}(47z^2 + z^3)$ \\[1em]
Electric Field $\bm{E}(\bm{x})$: & $-\begin{bmatrix}
100x \\ 100y \\ 10z
\end{bmatrix}$ & $-\begin{bmatrix}
94x + 3x^2 \\ 94y + 3y^2 \\ \frac{47}{5}z + \frac{3}{10}z^2
\end{bmatrix}$ & $-\frac{1}{3}\begin{bmatrix}
100x^3 \\ 100y^3 \\ 10z^3
\end{bmatrix}$
\end{tabular}
\caption{Electric fields for 3D potential well problems with uniform magnetic field $\bm{B} = 100\,\hat{\bm{z}}$}\label{EM3D}
\end{table}

\subsection{Two Dimensional Model Results}\subsubsection{Results of Electric Potential Well Problems}
\begin{figure}[H]
\centering
\begin{tabular}{ccc}
Quadratic & Cubic & Quartic \\
\includegraphics[scale=0.325]{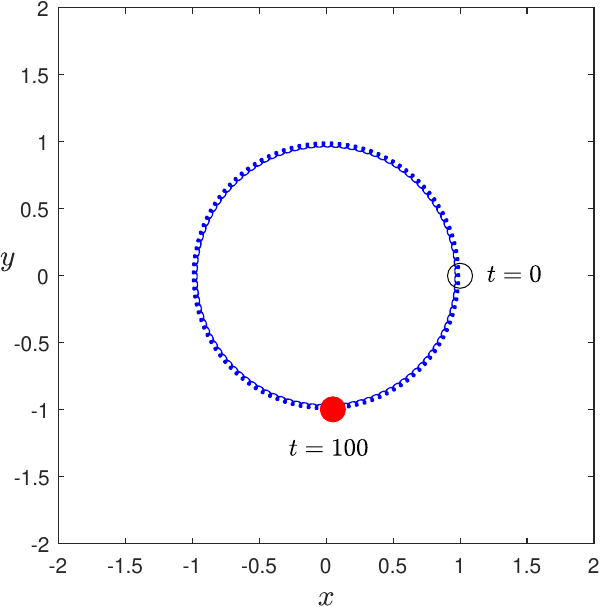} &
\includegraphics[scale=0.325]{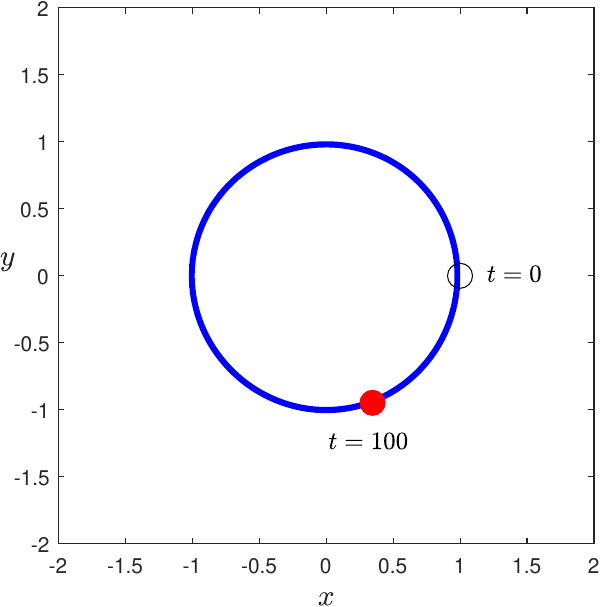} &
\includegraphics[scale=0.325]{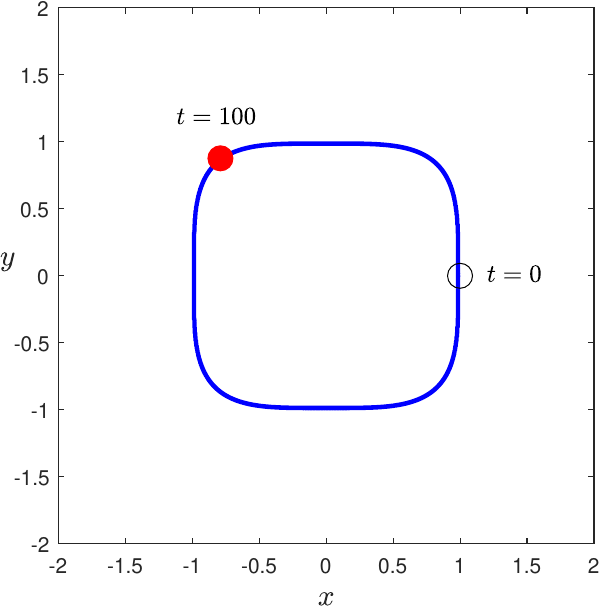} \\
\includegraphics[scale=0.325]{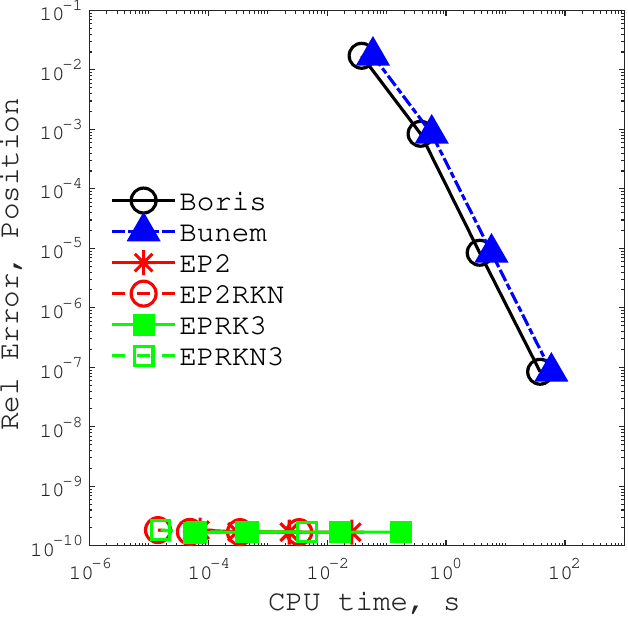} &
\includegraphics[scale=0.325]{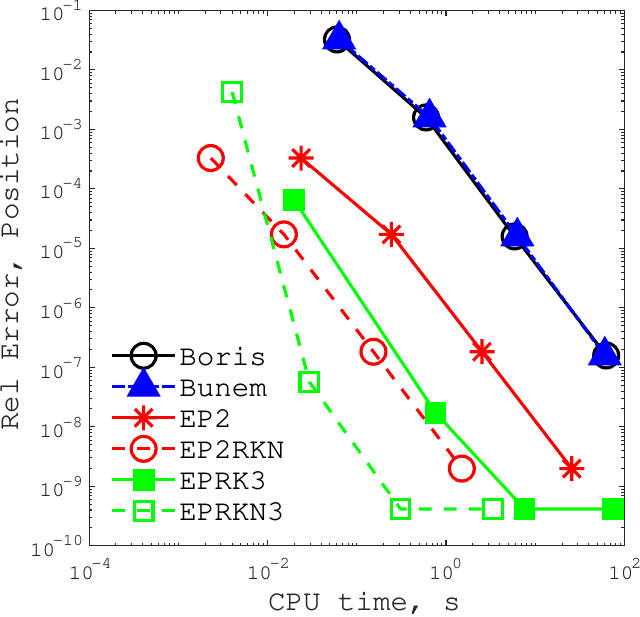} &
\includegraphics[scale=0.325]{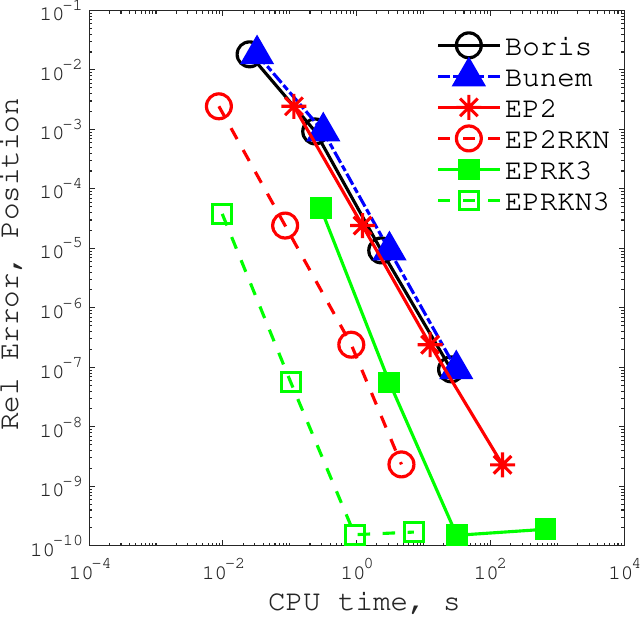}
\end{tabular}
\caption{Results for particle position, 2D potential well test problems with magnetic field $\bm{B} = 100\,\hat{\bm{z}}$: reference solution orbits (first row) and precision diagrams (second row). Boris/Buneman step sizes are $h = 10^{-3}$, $10^{-4}$, $10^{-5}$, $10^{-6}$ for the quadratic potential problem and $h = 10^{-4}$, $10^{-5}$, $10^{-6}$, $10^{-7}$ for the cubic/quartic potential problems. Exponential integrators step sizes are $h =$ 100, 10, 1, $10^{-1}$ for the quadratic potential problem and $h = 10^{-2}$, $10^{-3}$, $10^{-4}$, $10^{-5}$ for the cubic/quartic potential problems.}\label{EWell2D}
\end{figure}

\begin{figure}[H]
\centering
\begin{tabular}{ccc}
Quadratic & Cubic & Quartic \\
\includegraphics[scale=0.325]{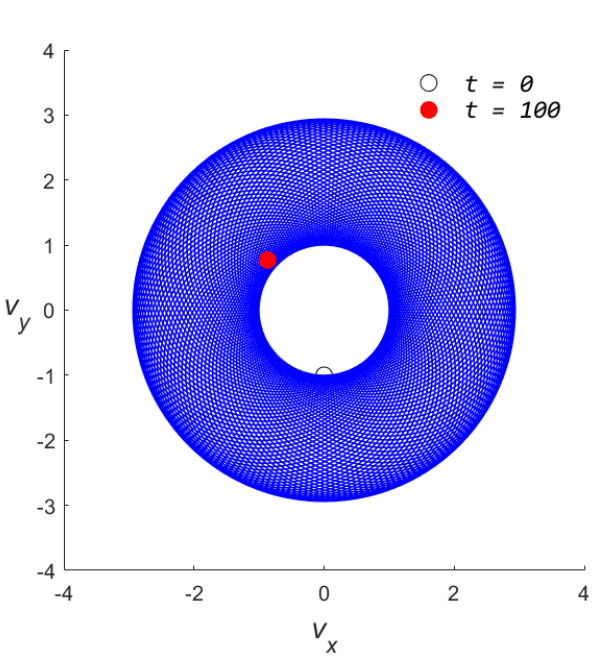} &
\includegraphics[scale=0.325]{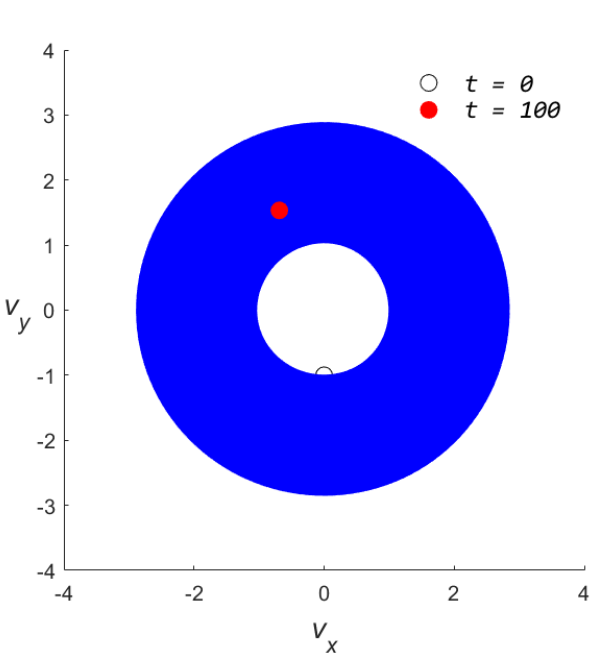} &
\includegraphics[scale=0.325]{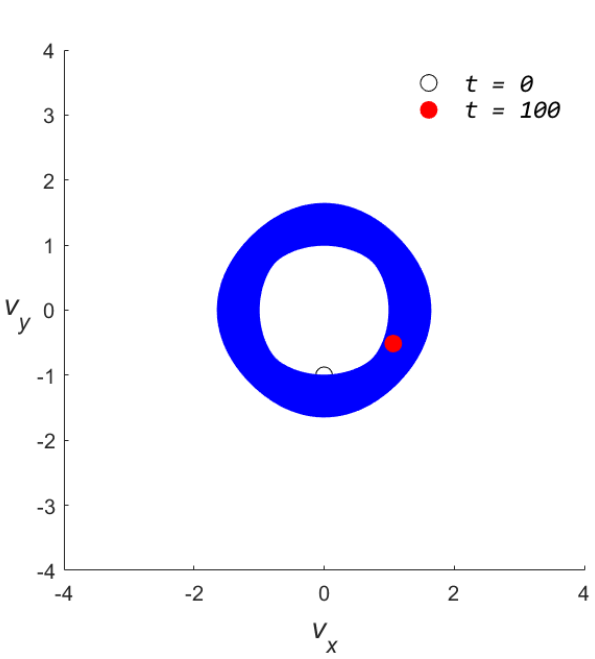} \\
\includegraphics[scale=0.325]{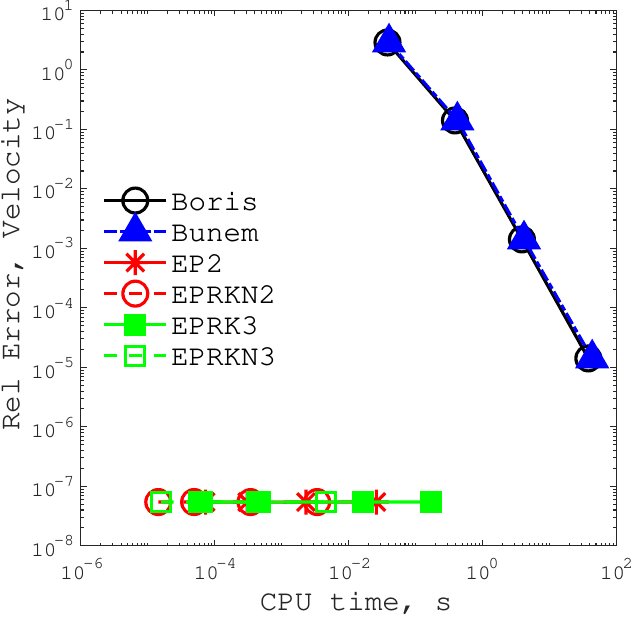} &
\includegraphics[scale=0.325]{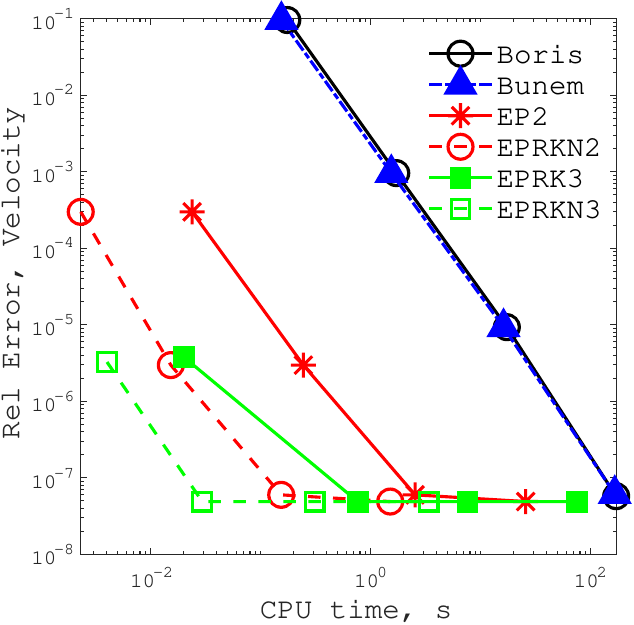} &
\includegraphics[scale=0.325]{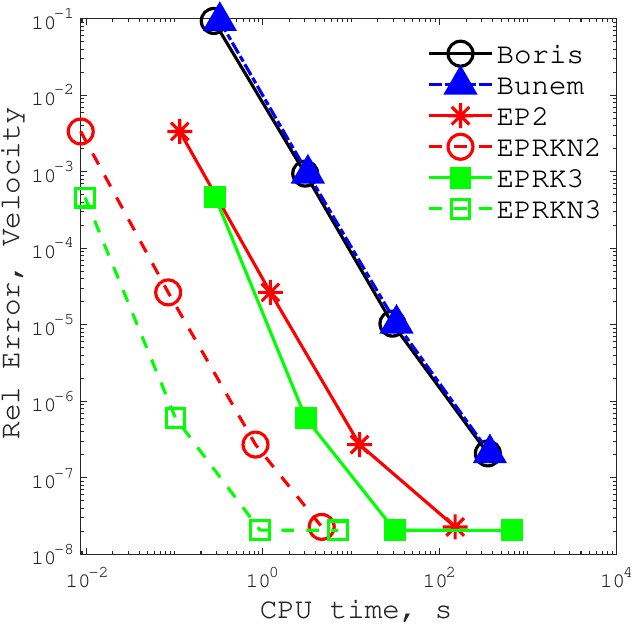}
\end{tabular}
\caption{Results for particle velocity, 2D potential well test problems with magnetic field $\bm{B} = 100\,\hat{\bm{z}}$: reference solution orbits (first row) and precision diagrams (second row). Boris/Buneman step sizes are $h = 10^{-3}$, $10^{-4}$, $10^{-5}$, $10^{-6}$ for the quadratic potential problem and $h = 10^{-4}$, $10^{-5}$, $10^{-6}$, $10^{-7}$ for the cubic/quartic potential problems. Exponential integrators step sizes are $h =$ 100, 10, 1, $10^{-1}$ for the quadratic potential problem and $h = 10^{-2}$, $10^{-3}$, $10^{-4}$, $10^{-5}$ for the cubic/quartic potential problems.}\label{EWellv2D}
\end{figure}

\begin{figure}[H]
\centering
\begin{tabular}{ccc}
Quadratic & Cubic & Quartic \\
\includegraphics[scale=0.325]{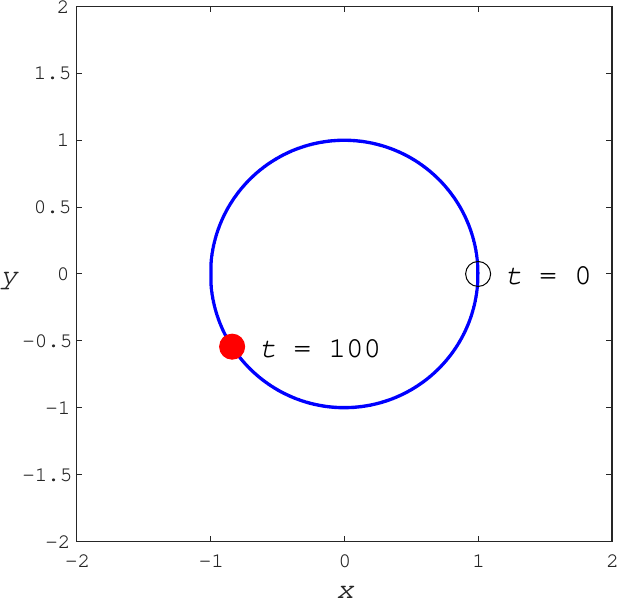} &
\includegraphics[scale=0.325]{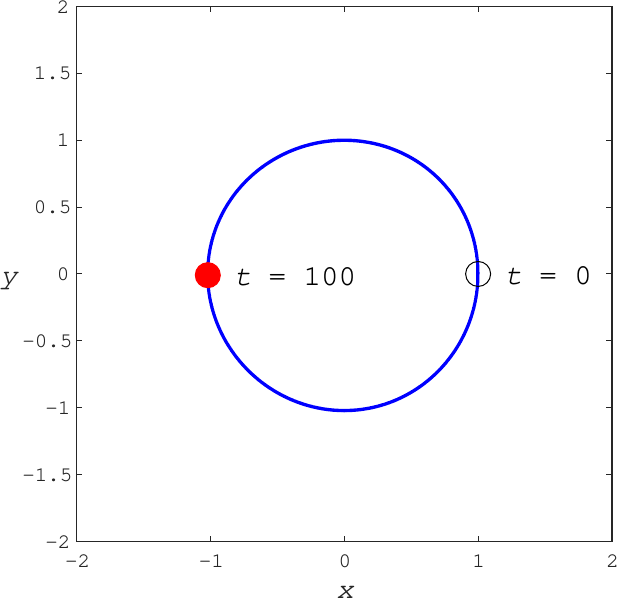} &
\includegraphics[scale=0.325]{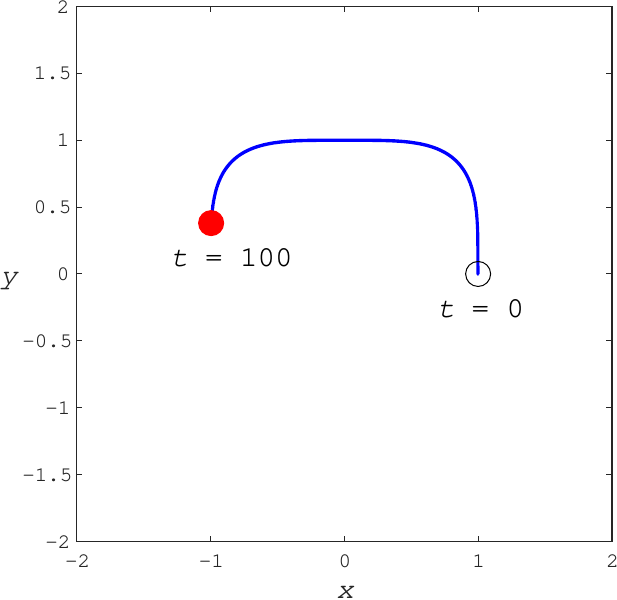} \\
\includegraphics[scale=0.325]{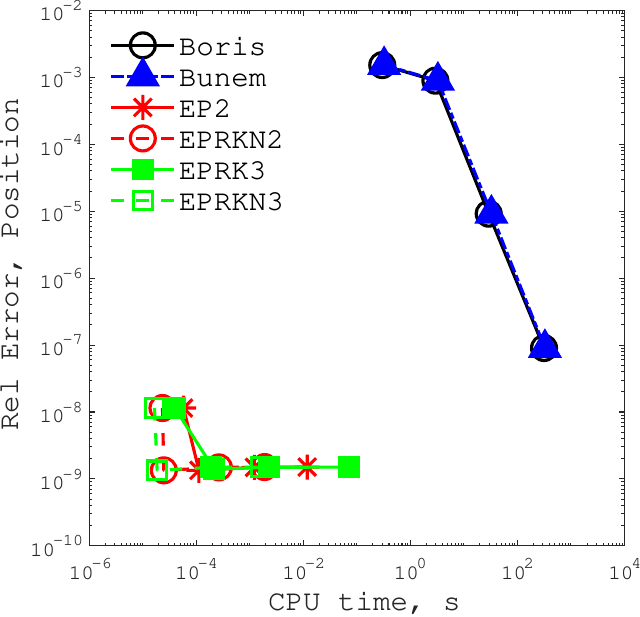} &
\includegraphics[scale=0.325]{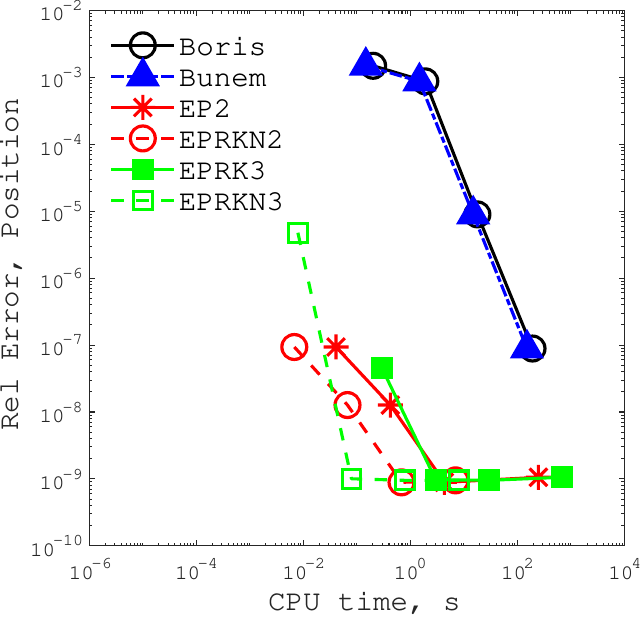} &
\includegraphics[scale=0.325]{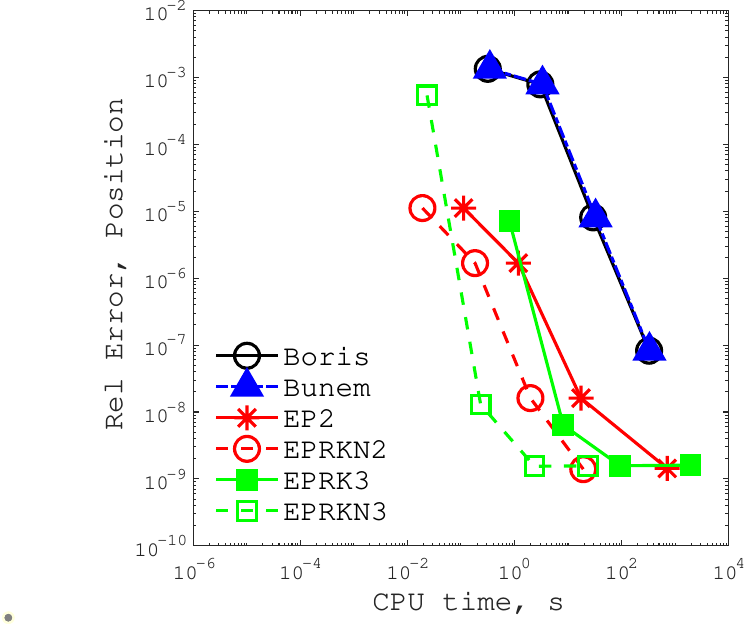}
\end{tabular}
\caption{Results for particle position, 2D potential well test problems with magnetic field $\bm{B} = 1000\,\hat{\bm{z}}$: reference solution orbits (first row) and precision diagrams (second row). Boris/Buneman step sizes are $h = 10^{-4}$, $10^{-5}$, $10^{-6}$, $10^{-7}$. Exponential integrators step sizes are $h =$ 100, 10, 1, $10^{-1}$ for the quadratic potential problem and $h = 10^{-2}$, $10^{-3}$, $10^{-4}$, $10^{-5}$ for the cubic/quartic potential problems.}\label{EWell2D_B1000}
\end{figure}

\begin{figure}[H]
\centering
\begin{tabular}{ccc}
Quadratic & Cubic & Quartic \\
\includegraphics[scale=0.325]{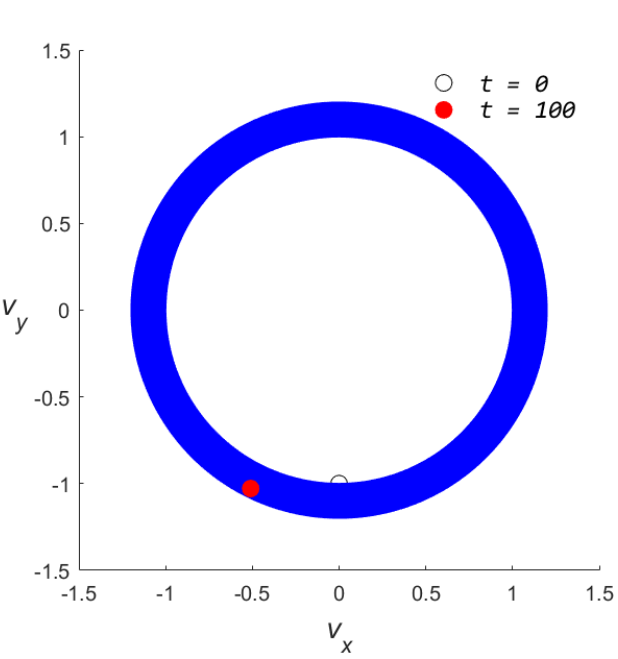} &
\includegraphics[scale=0.325]{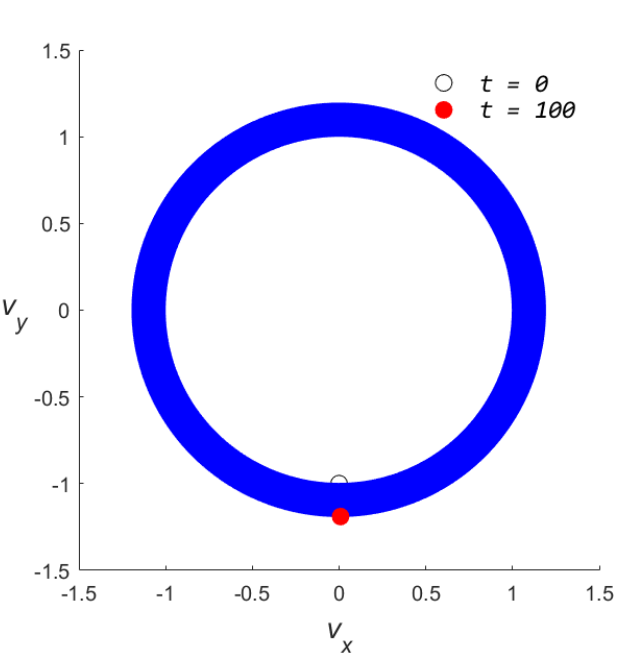} &
\includegraphics[scale=0.325]{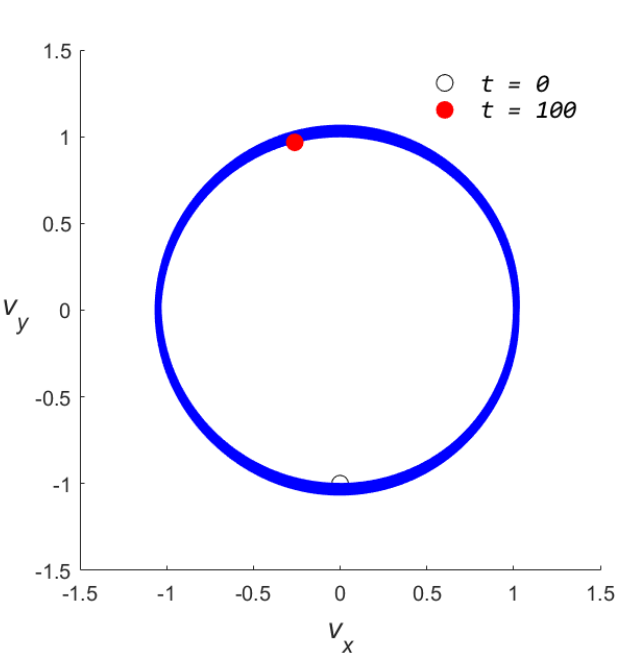} \\
\includegraphics[scale=0.325]{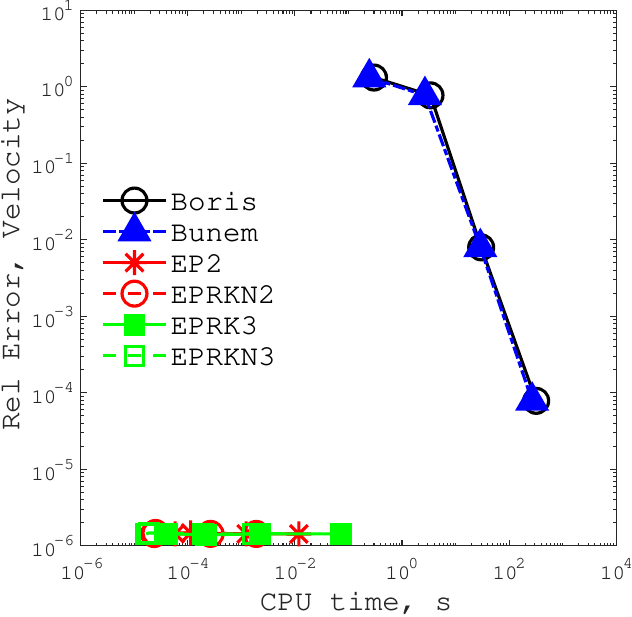} &
\includegraphics[scale=0.325]{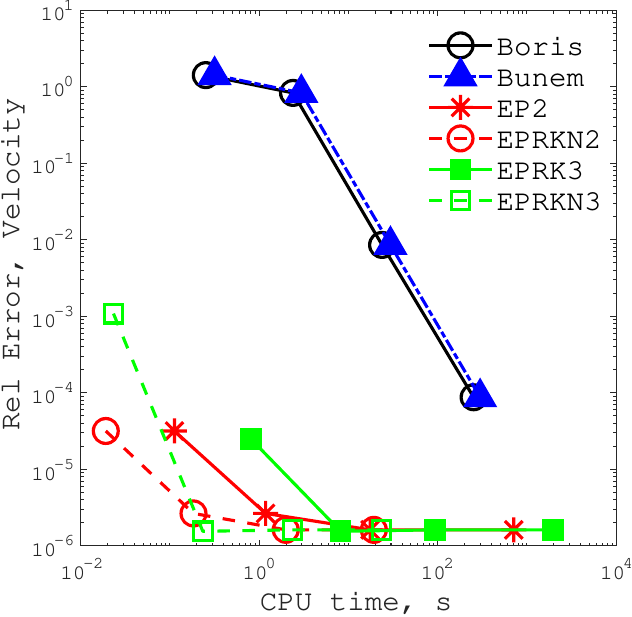} &
\includegraphics[scale=0.325]{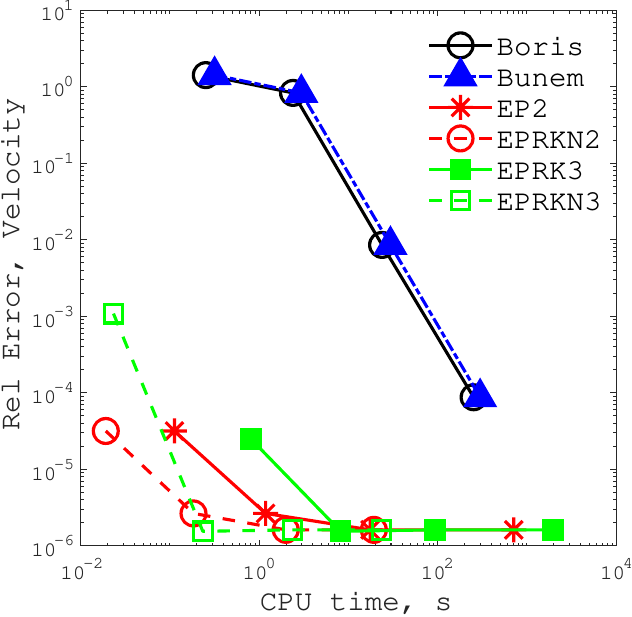}
\end{tabular}
\caption{Results for particle velocity, 2D potential well test problems with magnetic field $\bm{B} = 1000\,\hat{\bm{z}}$: reference solution orbits (first row) and precision diagrams (second row). Boris/Buneman step sizes are $h = 10^{-4}$, $10^{-5}$, $10^{-6}$, $10^{-7}$. Exponential integrators step sizes are $h =$ 100, 10, 1, $10^{-1}$ for the quadratic potential problem and $h = 10^{-2}$, $10^{-3}$, $10^{-4}$, $10^{-5}$ for the cubic/quartic potential problems.}\label{EWellv2D_B1000}
\end{figure}

Figures \ref{EWell2D} and \ref{EWellv2D} display plots of the experiment results for particle position and velocity, respectively, with magnetic field $\bm{B} = 100\,\hat{\bm{z}}$. Figures \ref{EWell2D_B1000} and \ref{EWellv2D_B1000} shows results for particle position and velocity, respectively, for the same experiment but with the stronger magnetic field $\bm{B} = 1000\,\hat{\bm{z}}$. For each figure, plots of the reference solution are in the top row. Work-precision diagrams are in the bottom row. Results for the quadratic, cubic, and quartic potential problems are in the left, center, and right columns, respectively.

Note that for the quadratic potential well problem the exponential integrators exhibit superior performance as expected, since this is a linear problem. For the cubic potential well problem, all of the exponential integrators compute solutions more accurately and efficiently than the Boris and Buneman pushers. For the quartic potential well problem, the exponential integrators are at least competitive if not better than the Boris and Buneman particle pushers. In particular, the Nyström-type exponential integrators consistently outperform both standard exponential methods and the Boris and Buneman pushers for all levels of accuracy yielding significant improvements in efficiency. Additionally, experimental results with the stronger magnetic field $\bm{B} = 1000\,\hat{\bm{z}}$ show that the exponential integrators significantly outperform the conventional Boris and Buneman pushers in terms of accuracy. Tables \ref{CPUtratio_Ewell_2D} and \ref{CPUtratio_Ewell_B1000} quantify these gains in computational efficiency by showing the average ratios of the CPU times of the standard exponential integrators to the Nyström-type exponential integrators for each test problem for magnetic fields $\bm{B} = 100\,\hat{\bm{z}}$ and $\bm{B} = 1000\,\hat{\bm{z}}$, respectively.

\begin{table}[H]
\centering
\begin{tabular}{lccc}
& Quadratic Well & Cubic Well & Quartic Well \\[0.5em]
EP2/EPRKN2 & 7.56 & 16.97 & 29.50 \\[0.5em]
EPRK3/EPRKN3 & 36.86 & 22.46 & 84.73
\end{tabular}
\caption{Average CPU time ratios of standard exponential integrators to Nyström-type exponential integrators for 2D potential well problems.}\label{CPUtratio_Ewell_2D}
\end{table}

\begin{table}[H]
\centering
\begin{tabular}{lccc}
& Quadratic Well & Cubic Well & Quartic Well \\[0.5em]
EP2/EPRKN2 & 6.00 & 32.50 & 34.07 \\[0.5em]
EPRK3/EPRKN3 & 39.53 & 79.82 & 78.31
\end{tabular}
\caption{Average CPU time ratios of standard exponential integrators to Nyström-type exponential integrators for 2D potential well problems with magnetic field $\bm{B} = 1000\,\hat{\bm{z}}$.}\label{CPUtratio_Ewell_B1000}
\end{table}

\newpage
\subsubsection{Results of Gyroradius Problem}
\begin{figure}[H]
\centering
\begin{tabular}{cc}
Boris & Buneman \\
\includegraphics[scale=0.35]{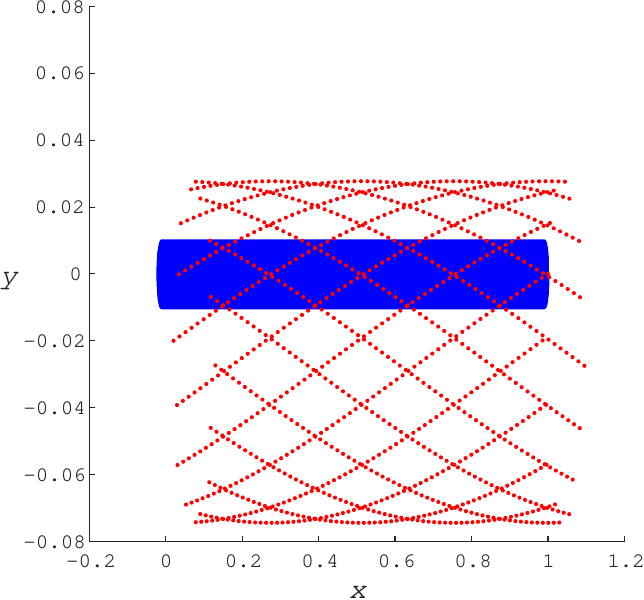} & \includegraphics[scale=0.35]{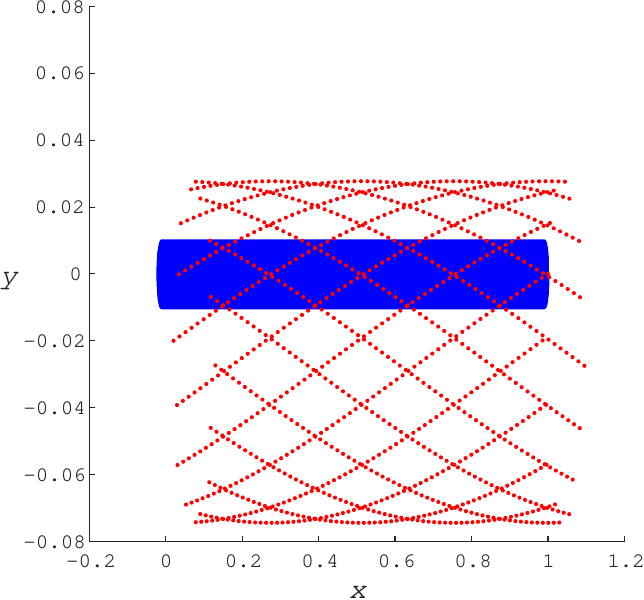} \\[1em]
EP2 & EPRK3 \\
\includegraphics[scale=0.35]{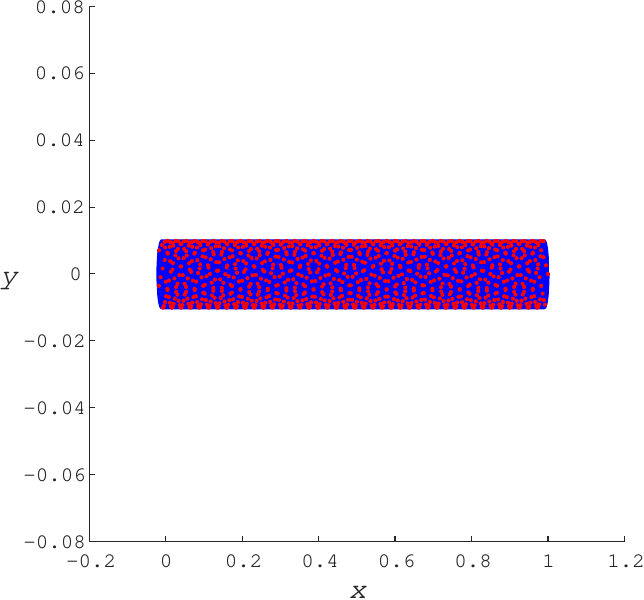} & \includegraphics[scale=0.35]{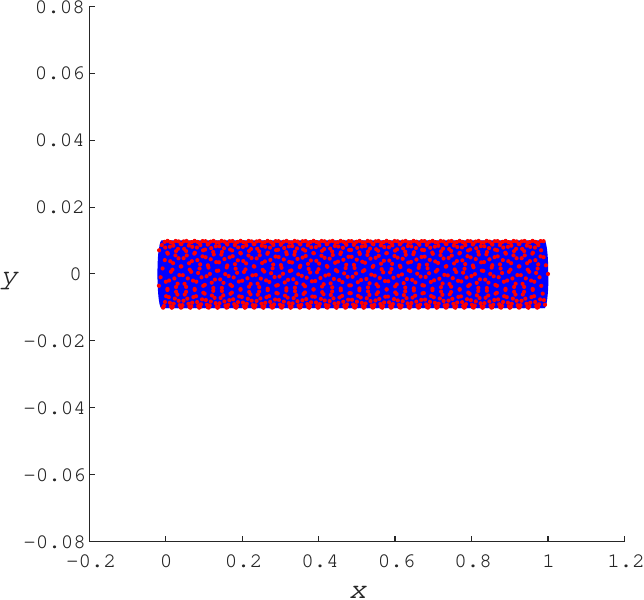} \\[1em]
EPRKN2 & EPRKN3 \\
\includegraphics[scale=0.35]{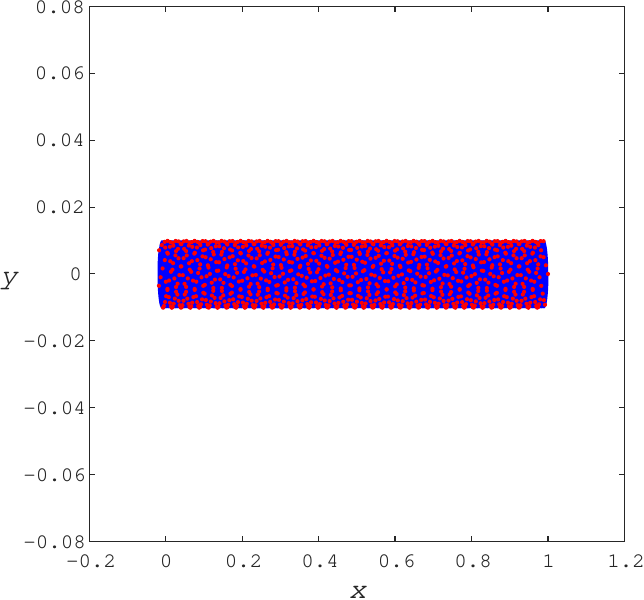} & \includegraphics[scale=0.35]{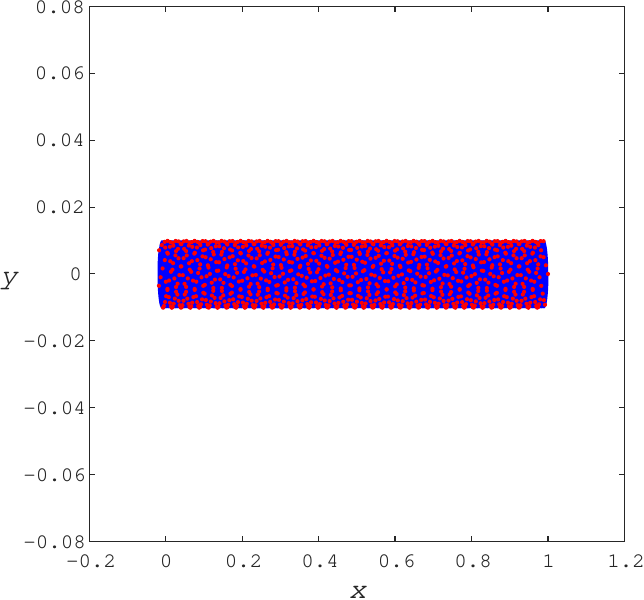}
\end{tabular}
\caption{Plots of computed trajectories for the gyroradius $\bm{E}\times\bm{B}$ drift problem. Solutions for step size $h = 0.001$ are solid blue and solutions for step size $h = 0.01$ are dotted red.}\label{gyroradius_ExB}
\end{figure}

Figure \ref{gyroradius_ExB} shows the experiment results for the gyroradius problem. For the "small" step size $h = 0.001$, all particle pushers compute the gyroradius accurately. However for the "large" step size $h = 0.1$, both the Boris and Buneman algorithms compute an artificially enlarged gyroradius while all the exponential integrators compute the correct gyroradius.

\newpage
\subsubsection{Results of Grad-$B$ Drift Problem}
\begin{figure}[H]
\begin{tabular}{ccc}
$\delta B = 0.1$ & $\delta B = 1$ & $\delta B = 10$ \\[0.5em]
\includegraphics[scale=0.325]{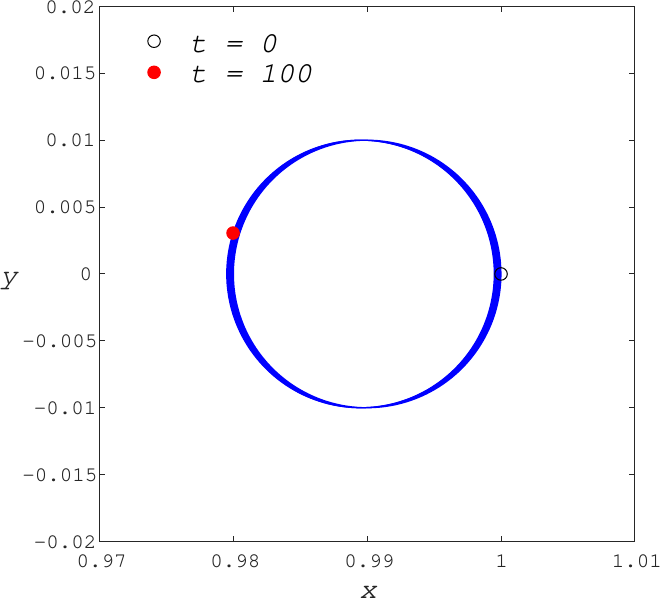} & \includegraphics[scale=0.325]{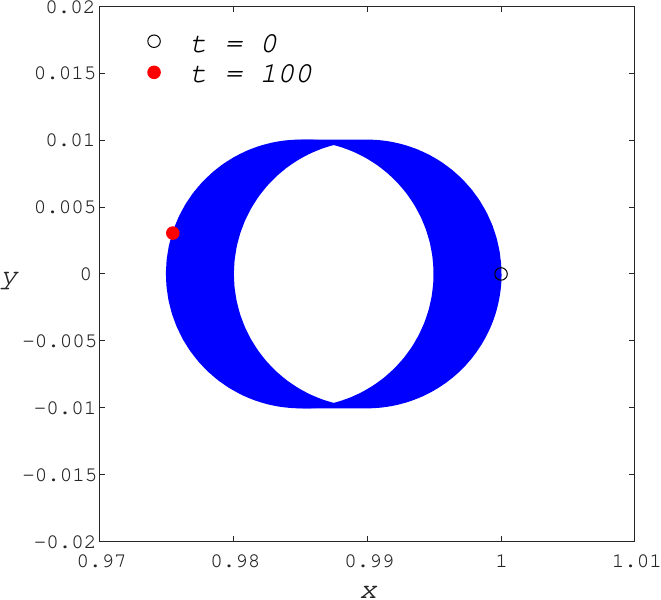} & \includegraphics[scale=0.325]{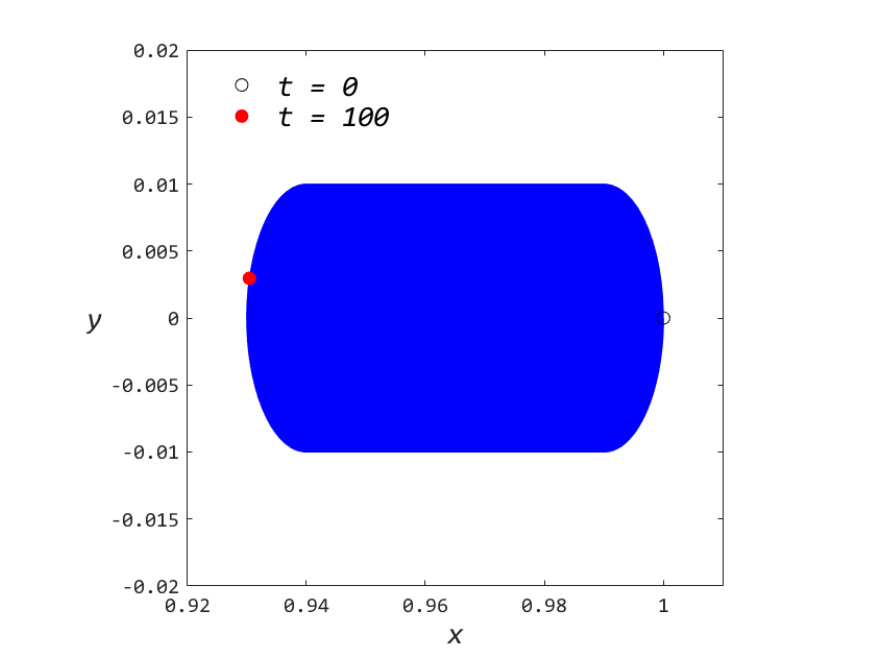} \\
\includegraphics[scale=0.325]{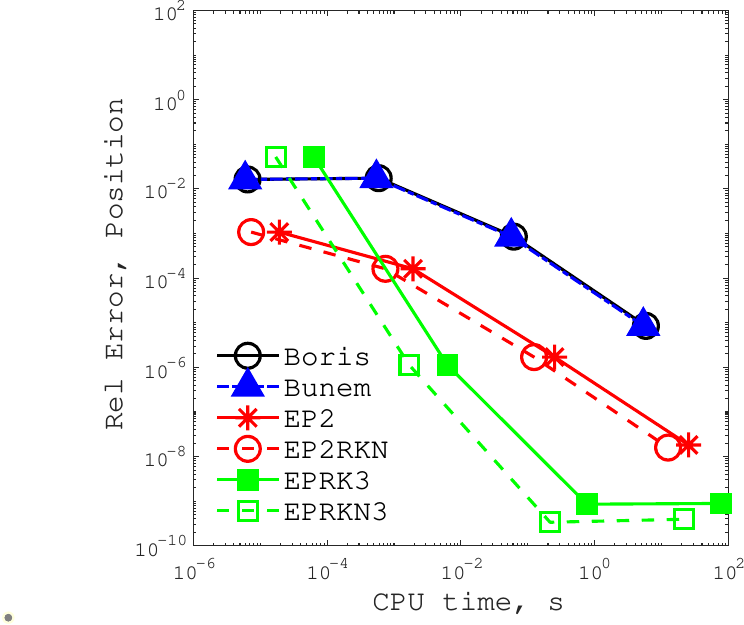} & \includegraphics[scale=0.325]{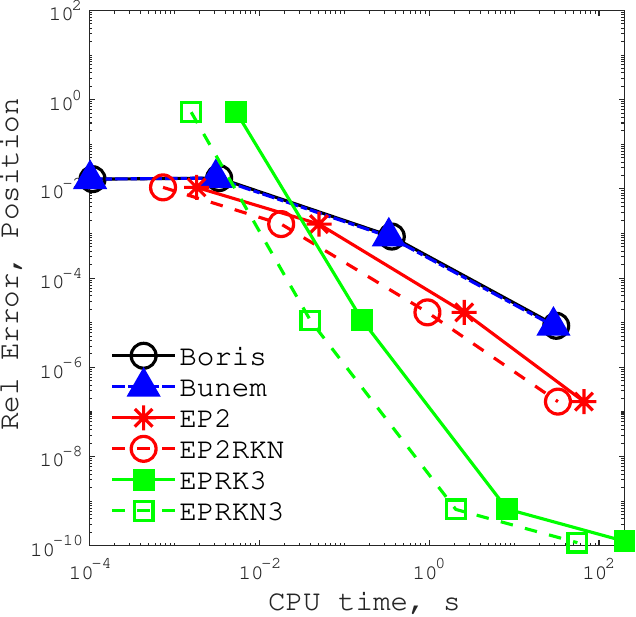} & \includegraphics[scale=0.325]{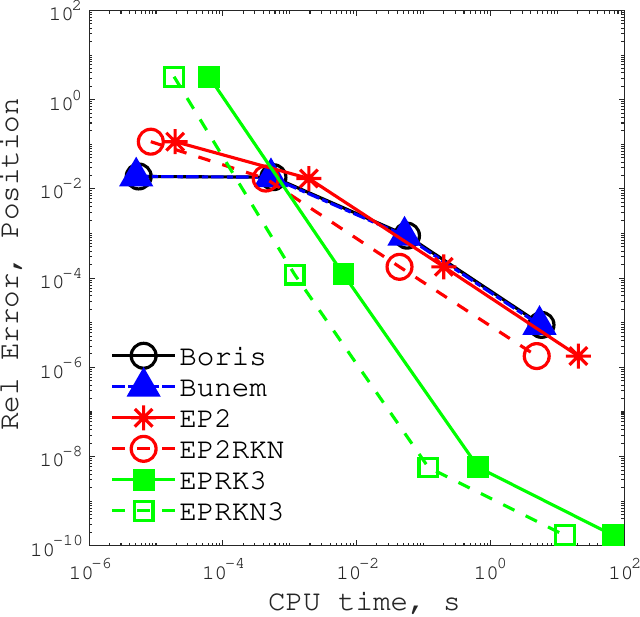} \\
\end{tabular}
\caption{Results for particle position, grad-$B$ drift problem: reference solution orbits (top row), and precision diagrams (bottom row). Boris/Buneman step sizes are $h = 10^{-2}$, $10^{-3}$, $10^{-4}$, $10^{-5}$. EP2/EPRK3 step sizes are $h = 10^{-1}$, $10^{-2}$, $10^{-3}$, $10^{-4}$.}\label{GradBDrift}
\end{figure}

\begin{figure}[H]
\begin{tabular}{ccc}
$\delta B = 0.1$ & $\delta B = 1$ & $\delta B = 10$ \\[0.5em]
\includegraphics[scale=0.325]{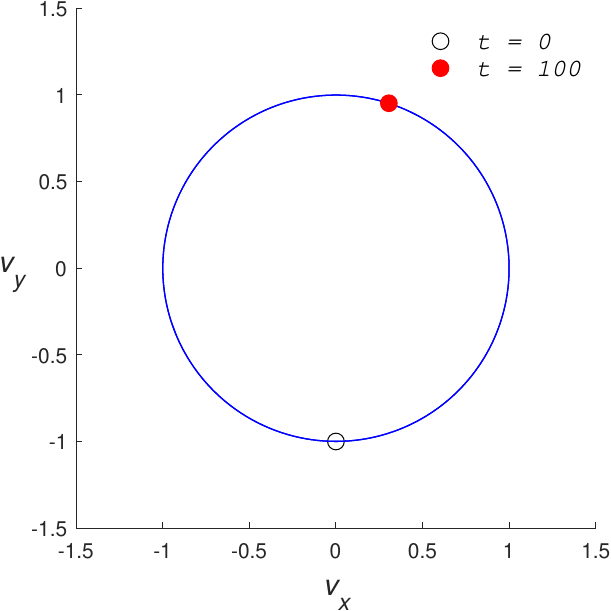} & \includegraphics[scale=0.325]{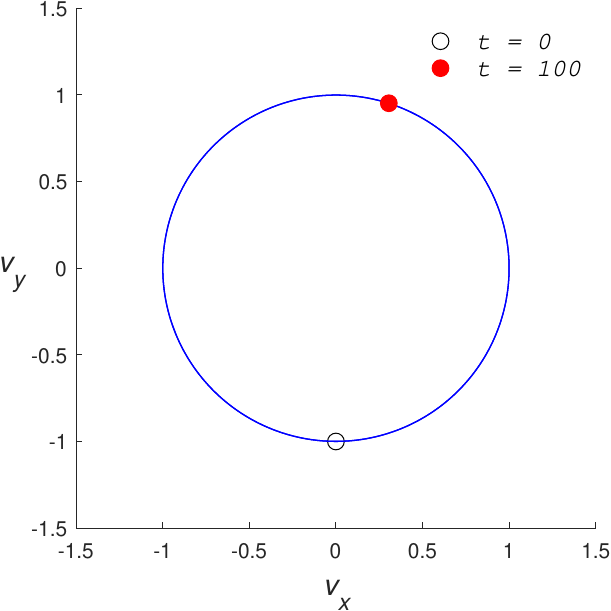} & \includegraphics[scale=0.325]{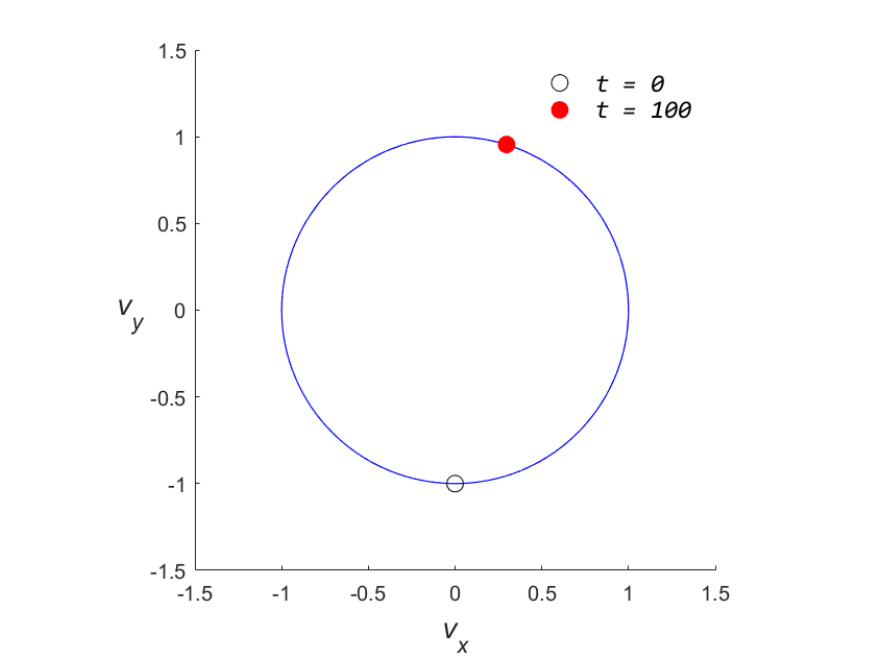} \\
\includegraphics[scale=0.325]{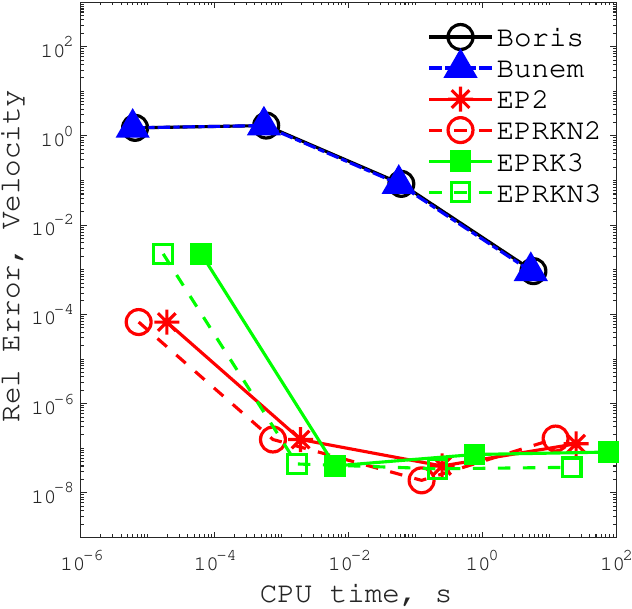} & \includegraphics[scale=0.325]{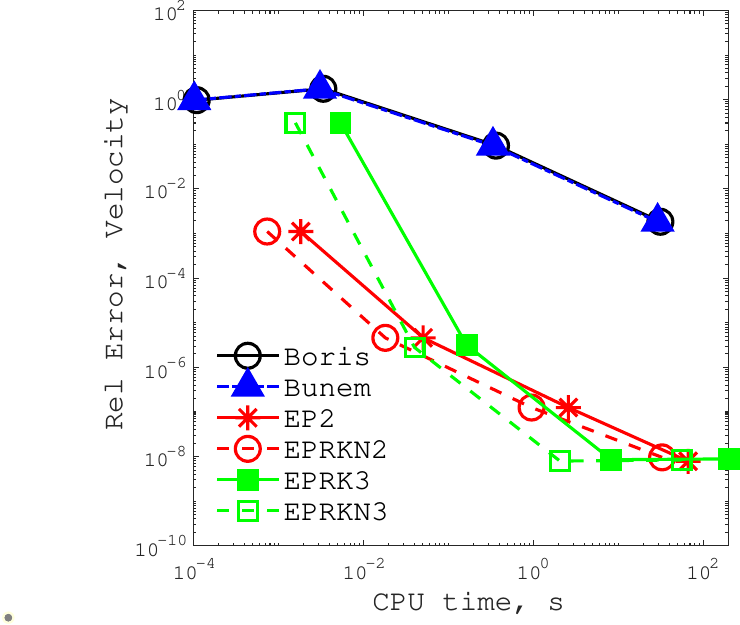} & \includegraphics[scale=0.325]{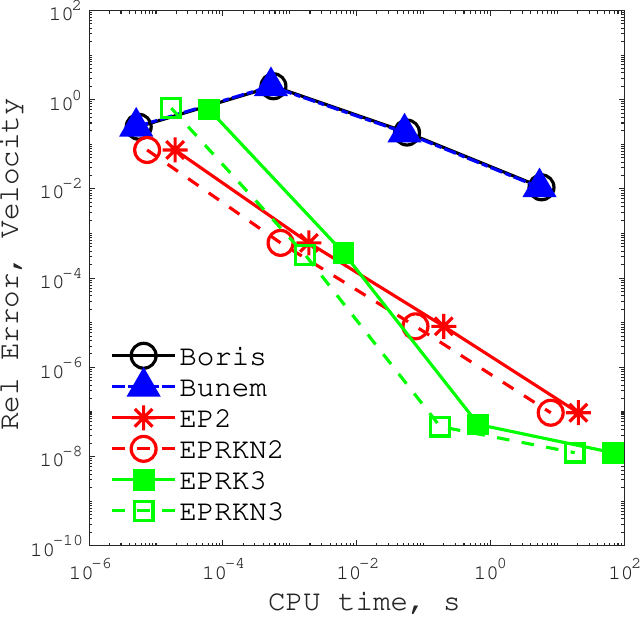} \\
\end{tabular}
\caption{Results for particle velocity, grad-$B$ drift problem: reference solution orbits (top row), and precision diagrams (bottom row). Boris/Buneman step sizes are $h = 10^{-2}$, $10^{-3}$, $10^{-4}$, $10^{-5}$. EP2/EPRK3 step sizes are $h = 10^{-1}$, $10^{-2}$, $10^{-3}$, $10^{-4}$.}\label{GradBDriftv}
\end{figure}

Figure \ref{GradBDrift} shows plots of the particle position reference solution orbits and the precision diagrams for $\delta B$ = 0.1, 1, 10. Figure \ref{GradBDriftv} shows plots of the particle velocity reference solution orbits and the precision diagrams for $\delta B$ = 0.1, 1, 10. Again the Nyström-type exponential integrators compute much faster than the standard exponential integrators. Table \ref{CPUtratio_gradB} shows the average ratios of the CPU times of the standard exponential integrators to the Nyström-type exponential integrators.

\begin{table}[H]
\centering
\begin{tabular}{lccc}
& $\delta B = 0.1$ & $\delta B = 1$ & $\delta B = 10$ \\[0.5em]
EP2/EPRKN2 & 2.03 & 2.04 & 4.19 \\[0.5em]
EPRK3/EPRKN3 & 3.57 & 3.60 & 5.24
\end{tabular}
\caption{Average CPU time ratios of standard exponential integrators to Nyström-type exponential integrators for the grad-$B$ drift problems.}\label{CPUtratio_gradB}
\end{table}

\subsection{Three Dimensional Model Results}
\begin{figure}[H]
\centering
\begin{tabular}{ccc}
Quadratic & Cubic & Quartic \\
\includegraphics[scale=0.3]{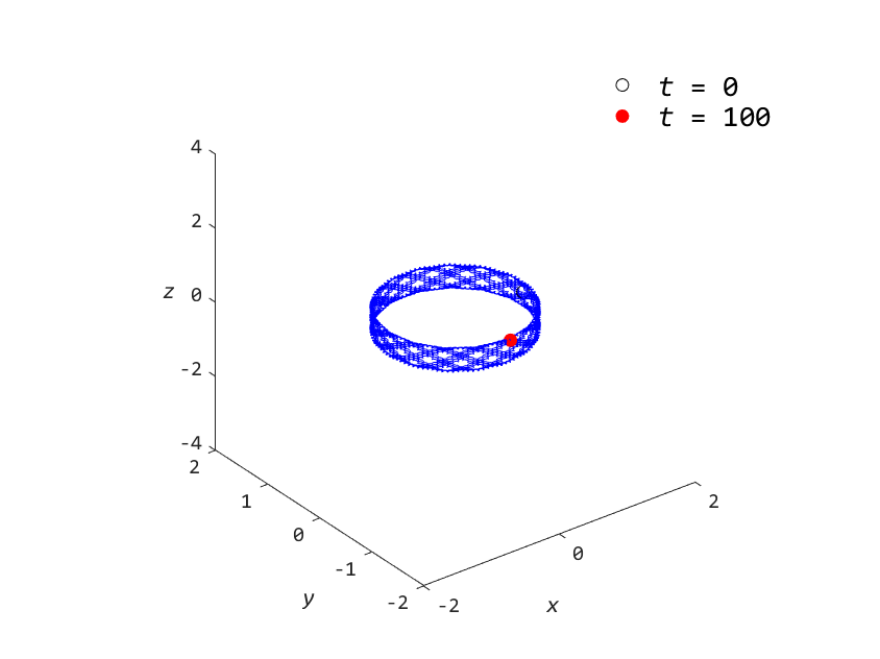} &
\includegraphics[scale=0.3]{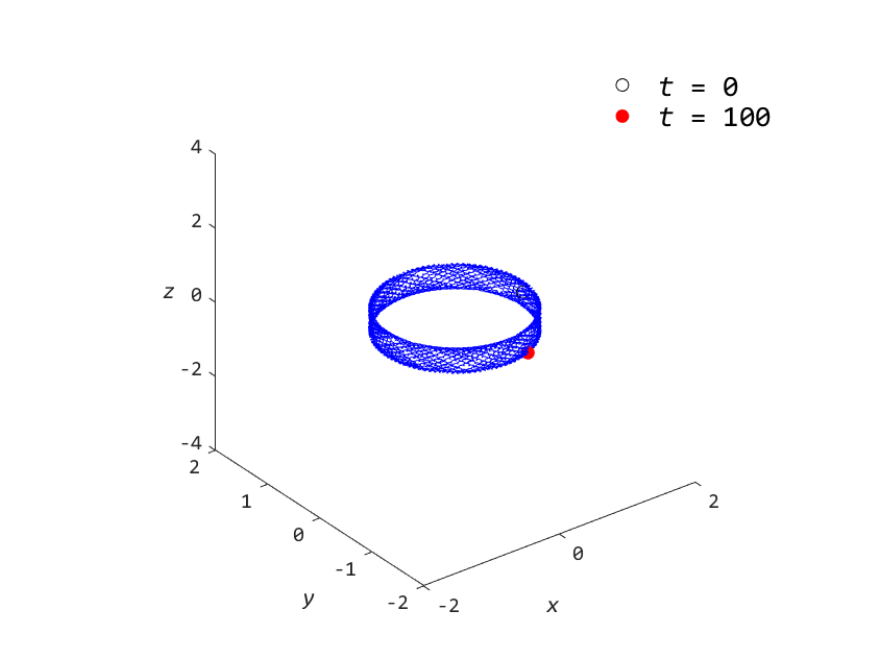} &
\includegraphics[scale=0.3]{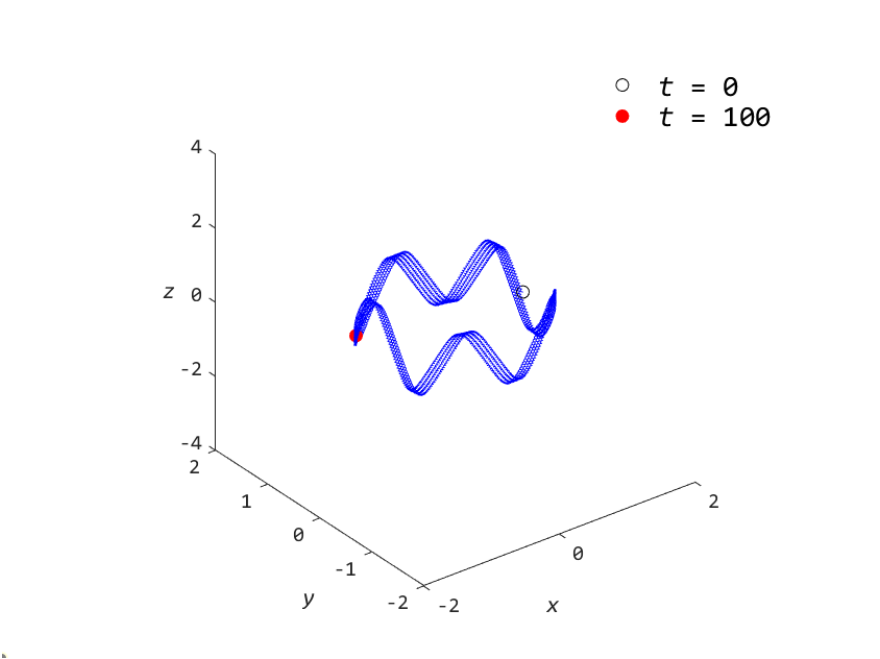} \\
\includegraphics[scale=0.3]{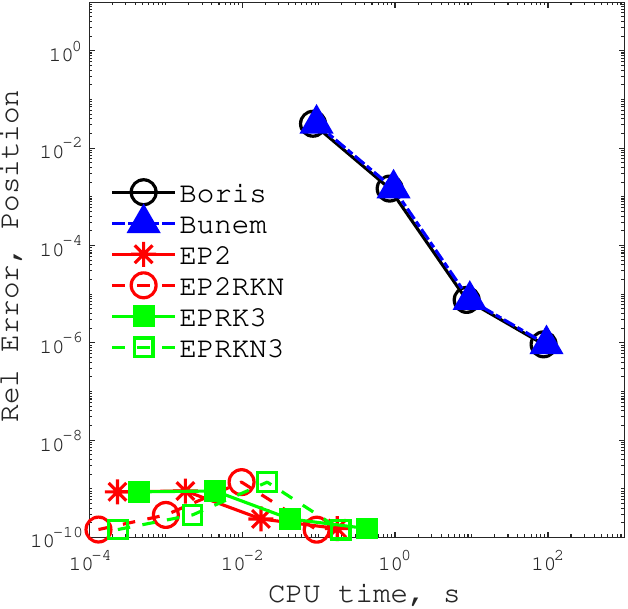} &
\includegraphics[scale=0.3]{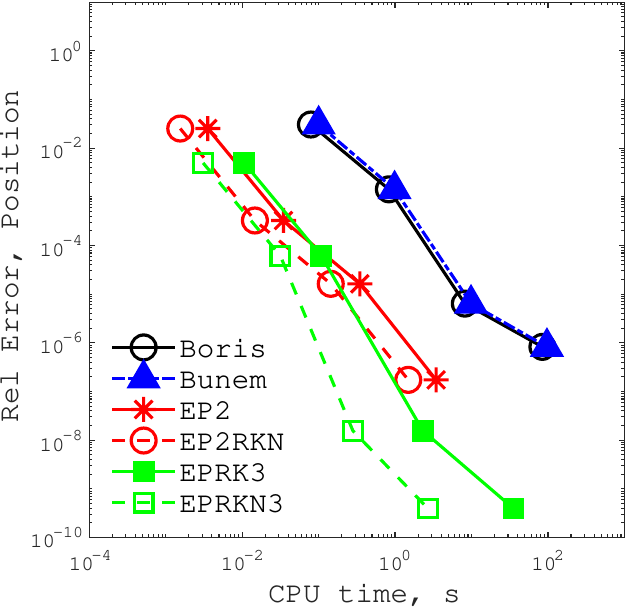} &
\includegraphics[scale=0.3]{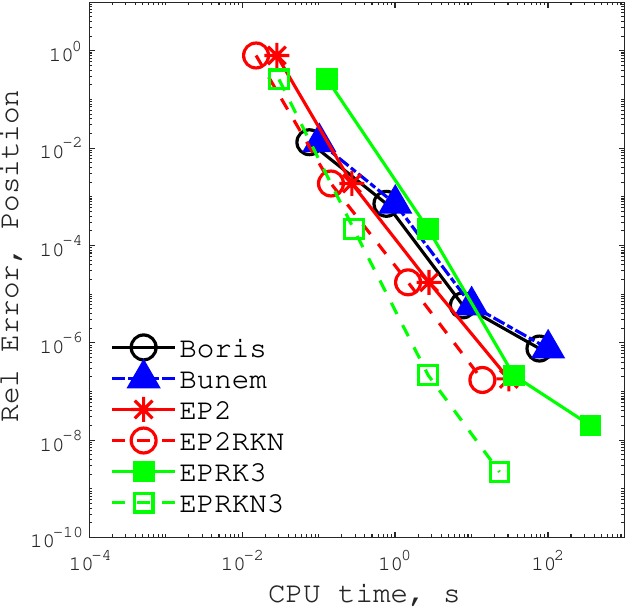}
\end{tabular}
\caption{Results for particle position, 3D potential well test problems: reference solution orbits (first row) and precision diagrams (second row). Boris/Buneman step sizes are $h = 10^{-2}$, $10^{-3}$, $10^{-4}$, $10^{-5}$ for the quadratic well problem, $h = 10^{-3}$, $10^{-4}$, $10^{-5}$, $10^{-6}$ for the cubic/quartic potential problems. Exponential integrators step sizes are $h =$ 100, 10, 1, $10^{-1}$ for the quadratic potential problem and $h = 10^{-1}$, $10^{-2}$, $10^{-3}$, $10^{-4}$ for the cubic/quartic potential problems.}\label{EWell3D}
\end{figure}

\begin{figure}[H]
\centering
\begin{tabular}{ccc}
Quadratic & Cubic & Quartic \\
\includegraphics[scale=0.3]{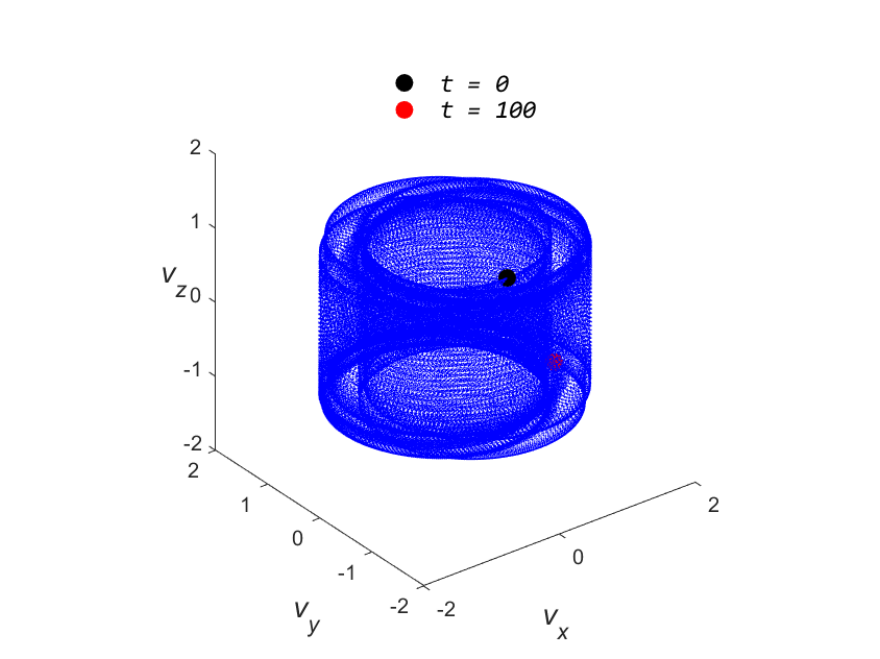} &
\includegraphics[scale=0.3]{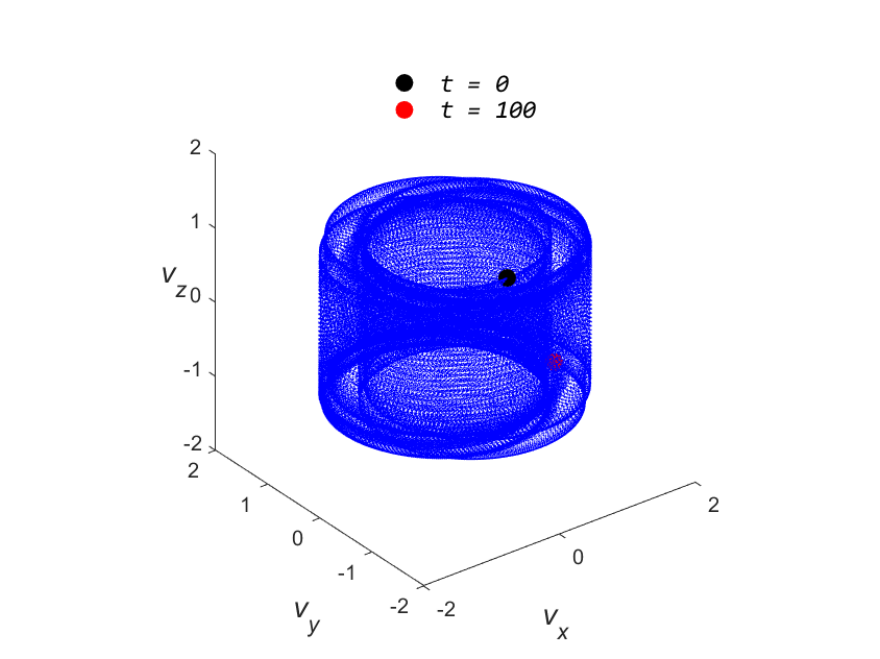} &
\includegraphics[scale=0.3]{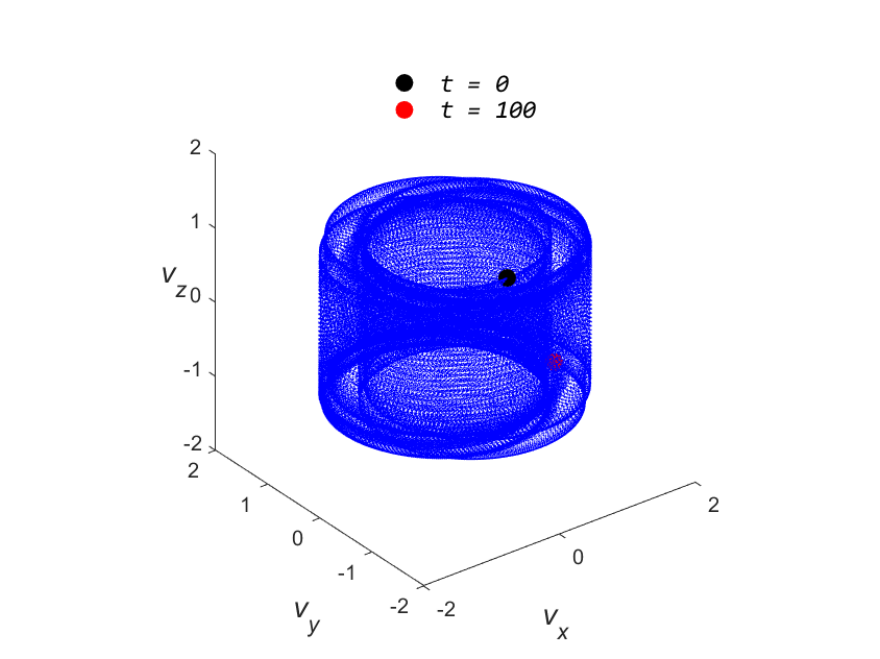} \\
\includegraphics[scale=0.3]{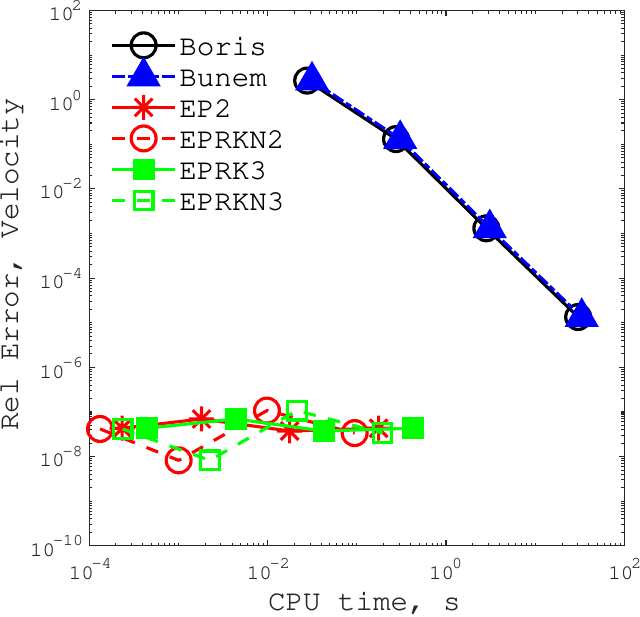} &
\includegraphics[scale=0.3]{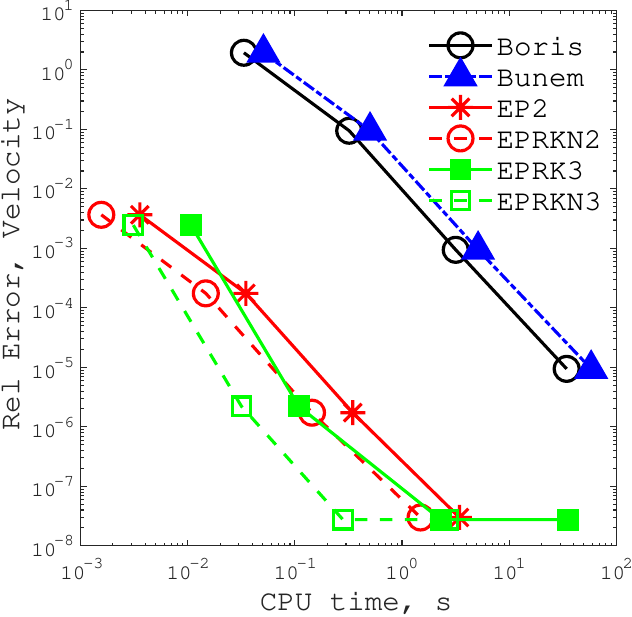} &
\includegraphics[scale=0.3]{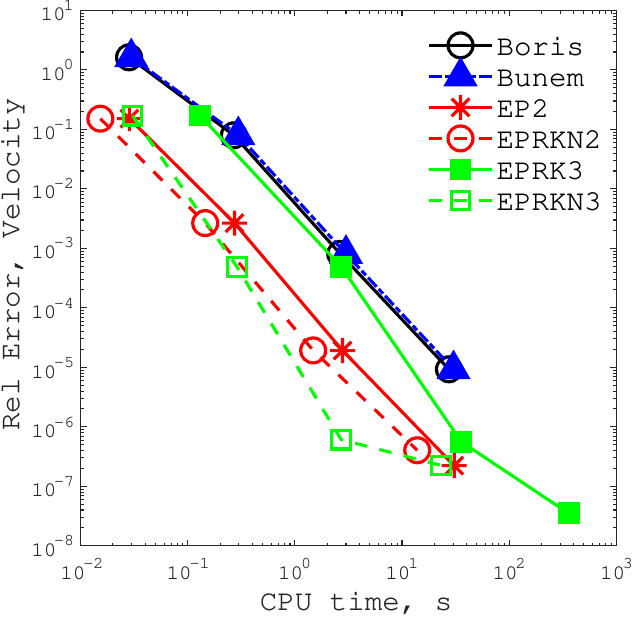}
\end{tabular}
\caption{Results for particle velocity, 3D potential well test problems: reference solution orbits (first row) and precision diagrams (second row). Boris/Buneman step sizes are $h = 10^{-2}$, $10^{-3}$, $10^{-4}$, $10^{-5}$ for the quadratic well problem, $h = 10^{-3}$, $10^{-4}$, $10^{-5}$, $10^{-6}$ for the cubic/quartic potential problems. Exponential integrators step sizes are $h =$ 100, 10, 1, $10^{-1}$ for the quadratic potential problem and $h = 10^{-1}$, $10^{-2}$, $10^{-3}$, $10^{-4}$ for the cubic/quartic potential problems.}\label{EWellv3D}
\end{figure}

Figures \ref{EWell3D} and \ref{EWellv3D} show plots of the reference solution orbits and the work-precision diagrams for particle position and velocity, respectively, for the three dimensional electric potential well problems. As expected, all the exponential integrators exhibit superior performance for the quadratic potential well problem. For the cubic potential well problem, the exponential integrators outperform the Boris and Buneman algorithms in terms of computation speed for comparable levels of accuracy. For the quartic potential well problem, the exponential integrators are competitive with the Boris and Buneman algorithms. For all of the test problems, the Nyström-type exponential integrators compute faster than the standard exponential integrators and outperform the Boris and Buneman integrators. The average CPU time ratios of the standard exponential integrators to the Nyström-type exponential integrators are shown in table \ref{CPUtratio_Ewell_3D}.

\begin{table}[H]
\centering
\begin{tabular}{lccc}
& Quadratic Well & Cubic Well & Quartic Well \\[0.5em]
EP2/EPRKN2 & 1.85 & 2.32 & 2.18 \\[0.5em]
EPRK3/EPRKN3 & 2.18 & 12.57 & 15.16
\end{tabular}
\caption{Average CPU time ratios of standard exponential integrators to Nyström-type exponential integrators for 3D potential well problems.}\label{CPUtratio_Ewell_3D}
\end{table}

\section{Conclusions and Future Work}
\subsection{Summary}\label{c3_sec:Summary}
In this study we derived Nyström-type exponential integrators induced by partitioning standard exponential methods. In particular, we partitioned the second-order EP2 and the third-order EPRK3 methods corresponding to the $\bm{x}$ and $\bm{v}$ components to construct second-order EPRKN2 and the third-order EPRKN3 methods that effectively solve the particle pushing problem as a second-order differential equation. These Nyström-type exponential integrators exploit the mathematical structure of the Newtonian formulation of the particle pushing problem to improve computational efficiency. Numerical experiments demonstrate that the Nyström exponential integrators exhibit significant improvements in computation times compared to the standard exponential integrators for the same level of accuracy for both the two dimensional and three dimensional models. This work shows that Nyström exponential integrators are a promising alternative to solve strongly magnetized particle pushing problems.

\subsection{Future Work}
While the exponential integrators we constructed offer improvements in the accuracy of the solution, they are not specifically designed to preserve any geometric properties of the solution exactly. In our future work we will investigate whether exponential schemes that preserve phase space volume or energy can be constructed. Such volume- or energy-preserving methods are desired when integration over very long time intervals has to be done. Additionally, our numerical experiments showed that the computational savings offered by the new exponential methods are larger for linear or weakly nonlinear problems compared to strongly nonlinear configurations such as the quartic potential and the grad-$B$ drift problems. A possible approach we plan to investigate to address this issue is to develop a better quadrature for the nonlinear integral terms in the variation of constants Volterra integral equation which serves as the starting point for construction of an exponential integrator. We also plan to conduct numerical experiments with more complex electromagnetic field configurations for more realistic test problems and study their performance within PIC simulations. Finally, a thorough evaluation of these exponential integrators requires comparing them against the more advanced conventional particle pushers such as the modified Crank-Nicolson scheme \cite{Ricketson}, the filtered Boris algorithm \cite{Hairer1}, as well as volume-preserving methods based on operator splitting \cite{He}.


\newpage
\appendix
\section{Lagrange-Sylvester interpolation polynomial coefficients}
\label{coefficients}
This appendix describes the analytic expressions for the coefficients of the Lagrange-Sylvester interpolation polynomial for the test problems discussed in this work. Recall that the Jacobian matrix of the Newtonian form of the particle pushing problem is
\[
\bm{A} = \begin{bmatrix}
    \bm{O} & \bm{I} \\
    \bm{H} & \bm{\Omega}
\end{bmatrix},
\]
where $\bm{O}$ and $\bm{I}$ are the $d\times d$ zero and identity matrices, respectively, $\bm{H} = \partial\bm{f}_L/\partial\bm{x}$ is the Jacobian matrix of $\bm{f}_L$ with respect to $\bm{x}$, and $\bm{\Omega}$ is the $d\times d$ skew symmetrc matrix such that $\bm{\Omega\, v} = \frac{q}{m}\bm{v}\times\bm{B}$. Here, $d = 2$ for the two dimensional model, and $d = 3$ for the three dimensional model.

\subsection{Two Dimensional Model}
For the two dimensional model,
\[
\bm{H} = \frac{\partial\bm{f}_L}{\partial\bm{x}} =\begin{bmatrix}
H_{11} & H_{12} \\
H_{21} & H_{22}
\end{bmatrix}
\qquad\text{and}\qquad
\bm{\Omega} = \begin{bmatrix}
    \phantom{-}0 & \omega \\
    -\omega & 0
\end{bmatrix}, \quad\omega = \frac{qB}{m}.
\]
The characteristic polynomial of $\bm{A}$ is
\[
\det(z\bm{I}_{4\times 4} - \bm{A}) = z^4 + z^2 P + \lambda Q + R,
\]
where
\begin{align*}
    P & = \omega^2 - H_{11} - H_{22}, \\
    Q & = \omega(H_{12} - H_{21}), \\
    R & = H_{11}H_{22} - H_{12}H_{21}.
\end{align*}
Note that all particle pushing problems under consideration in this work are strongly magnetized, which implies $P \neq 0$.

\subsubsection{Electric Potential Well and Gyroradius Problems}
For these test problems $\bm{H}$ is a diagonal matrix, hence $Q = 0$ and the characteristic polynomial reduces to
\[
z^4 + z^2\,P + R.
\]
To determine the polynomial coefficients for the interpolation problem
\[
a_0 + a_1 z + a_2 z^2 + a_3 z^3 = \varphi_k(hz),
\]
there are several cases to consider.

If either $R = 0$ or $P^2 = 4R$, then the eigenvalues of $\bm{A}$ are
\[
z = 0, 0, \pm i\,\mu,
\]
where $\mu = \sqrt{P}$. For $\varphi_k = \varphi_1$ the polynomial coefficients are:
\begin{align*}
a_0 & = 1, \\[0.5em]
a_1 & = \dfrac{h}{2}, \\[0.5em]
a_2 & = \dfrac{1}{\mu^2}\left(1 - \dfrac{\sin(h\mu)}{h\mu}\right), \\[0.5em]
a_3 & = \dfrac{1}{\mu^3}\left(\dfrac{1}{2} - \dfrac{1 - \cos(h\mu)}{h\mu}\right).
\end{align*}
For $\varphi_k = \varphi_3$ the polynomial coefficients are:
\begin{align*}
a_0 & = \dfrac{1}{6}, \\[0.5em]
a_1 & = \dfrac{h}{24}, \\[0.5em]
a_2 & = \dfrac{1}{\mu^2}\left(\dfrac{1}{6} - S(h\mu)\right), \\[0.5em]
a_3 & = \dfrac{1}{\mu^3}\left(\dfrac{1}{24}h\mu - C(h\mu)\right),
\end{align*}
where
\[
S(z) = \frac{z - \sin(z)}{z^3}
\qquad\text{and}\qquad
C(z) = \frac{\cos(z) - 1 + \frac{1}{2}z^2}{z^3}.
\]

If $R \neq 0$, then the eigenvalues of $\bm{A}$ are
\[
z = \pm i\,\mu, \pm i\,\nu,
\]
where
\[
\mu = \sqrt{\frac{P + \sqrt{P^2 - 4R}}{2}} \qquad\text{and}\qquad \nu = \sqrt{\frac{P - \sqrt{P^2 - 4R}}{2}}.
\]
For $\varphi_k = \varphi_1$ the polynomial coefficients are:
\begin{align*}
a_0 & = \dfrac{1}{\mu^2 - \nu^2}\left(\mu^2\,\dfrac{\sin(h\nu)}{h\nu} - \nu^2\,\dfrac{\sin(h\mu)}{h\mu}\right), \\[0.5em]
a_1 & = \dfrac{1}{\mu^2 - \nu^2}\left(\dfrac{\mu^2}{\nu}\left(\dfrac{1 - \cos(h\nu)}{h\nu}\right) - \dfrac{\nu^2}{\mu}\left(\dfrac{1 - \cos(h\mu)}{h\mu}\right)\right), \\[0.5em]
a_2 & = \dfrac{1}{\mu^2 - \nu^2}\left(\dfrac{\sin(h\nu)}{h\nu} - \dfrac{\sin(h\mu)}{h\mu}\right), \\[0.5em]
a_3 & = \dfrac{1}{\mu^2 - \nu^2}\left(\dfrac{1}{\nu}\left(\dfrac{1 - \cos(h\nu)}{h\nu}\right) - \dfrac{1}{\mu}\left(\dfrac{1 - \cos(h\mu)}{h\mu}\right)\right).
\end{align*}
For $\varphi_k = \varphi_3$ the polynomial coefficients are:
\begin{align*}
a_0 & = \dfrac{1}{\mu^2 - \nu^2}\left(\mu^2 \, S(h\nu) - \nu^2 \, S(h\mu)\right), \\[0.5em]
a_1 & = \dfrac{1}{\mu^2 - \nu^2}\left(\dfrac{\mu^2}{\nu}\,C(h\nu) - \dfrac{\nu^2}{\mu}\,C(h\mu)\right), \\[0.5em]
a_2 & = \dfrac{1}{\mu^2 - \nu^2}\left(S(h\nu) - S(h\mu)\right), \\[0.5em]
a_3 & = \dfrac{1}{\mu^2 - \nu^2}\left(\dfrac{1}{\nu}\,C(h\nu) - \dfrac{1}{\mu}\,C(h\mu)\right).
\end{align*}

\subsubsection{Grad-\textit{B} Drift Problem}
For the grad-$B$ drift test problem, $R = 0$ and the characteristic polynomial reduces to
\[
z^4 + z^2\,P + z\,Q = z(z^3 + z\,P + Q).
\]
Hence, the eigenvalues of $\bm{A}$ are
\[
z = 0, \mu, \nu, \overline{\nu},
\]
where $\mu$ is the real root and the conjugate pair $\nu, \overline{\nu}$ are the complex roots of the cubic polynomial $z^3 + z\,P + Q$. The polynomial coefficients are thus:
\begin{align*}
    a_0 & = \varphi_k(0), \\[0.5em]
    a_1 & = \dfrac{|\nu|^4\varphi_k(h\mu)\mathrm{Im}(\nu) + \mu^2\mathrm{Im}(\overline{\nu}^3\varphi_k(h\nu)) + \mu^3\mathrm{Im}(\overline{\nu}^2\varphi_k(h\nu))}{\mu|\nu|^2\mathrm{Im}(\nu)(|\nu|^2 - 2\mu\,\mathrm{Re}(\nu) + \mu^2)}, \\[0.5em]
    a_2 & = -\dfrac{\mu\,\mathrm{Im}(\overline{\nu}^3\varphi_k(h\nu)) + 2|\nu|^2\varphi_k(h\mu)\mathrm{Re}(\nu)\mathrm{Im}(\nu) - \mu^3\mathrm{Im}(\overline{\nu}\varphi_k(h\nu))}{\mu|\nu|^2\mathrm{Im}(\nu)(|\nu|^2 - 2\mu\,\mathrm{Re}(\nu) + \mu^2)}, \\[0.5em]
    a_3 & = \dfrac{\mu\,\mathrm{Im}(\overline{\nu}^2\varphi_k(h\nu)) - \mu^2\mathrm{Im}(\overline{\nu}\varphi_k(h\nu)) + |\nu|^2\varphi_k(h\mu)\mathrm{Im}(\nu)}{\mu|\nu|^2\mathrm{Im}(\nu)(|\nu|^2 - 2\mu\,\mathrm{Re}(\nu) + \mu^2)},
\end{align*}
where $\mathrm{Re}(z)$ and $\mathrm{Im}(z)$ denote the real and imaginary parts of the complex argument $z$, respectively.

\subsection{Three Dimensional Model}
For the three dimensional electric potential well test problems, the block matrices $\bm{H}$ and $\bm{\Omega}$ are given by
\[
\bm{H} = \frac{\partial\bm{f}_L}{\partial\bm{x}} = \begin{bmatrix}
    H_{11} & 0 & 0 \\
    0 & H_{22} & 0 \\
    0 & 0 & H_{33}
\end{bmatrix}
\quad\text{and}\quad
\bm{\Omega} = \begin{bmatrix}
    \phantom{-}0 & \omega & 0 \\
    -\omega & 0 & 0 \\
    \phantom{-}0 & 0 & 0
\end{bmatrix}, \quad\omega = \frac{qB}{m}.
\]
The characteristic polynomial of $\bm{A}$ is
\[
\det(z\bm{I}_{6\times 6} - \bm{A}) = z^6 + z^4\,P + z^2\,R + T
\]
where
\begin{align*}
    P & = \omega^2 - H_{11} - H_{22} - H_{33}, \\
    R & = H_{11}H_{22} + H_{11}H_{33} + H_{22}H_{33} - \omega^2 H_{33}, \\
    T & = -H_{11}H_{22}H_{33}.
\end{align*}
Again we assume strongly magnetized particle pushing problems implying $P$ is always nonzero. To determine the polynomial coefficients, we next examine the various cases.

If $R = T = 0$, then the characteristic polynomial reduces to
\[
z^6 + z^4\,P = z^4(z^2 + P).
\]
Thus, the eigenvalues of $\bm{A}$ are
\[
z = 0, 0, 0, 0, \pm i\,\mu,
\]
where $\mu = \sqrt{P}$. Hence, for $\varphi_k = \varphi_1$ the polynomial coefficients are:
\begin{align*}
    a_0 & = 1, \\
    a_1 & = \frac{h}{2}, \\[0.5em]
    a_2 & = \frac{h^2}{6}, \\[0.5em]
    a_3 & = \frac{h^3}{24}, \\[0.5em]
    a_4 & = \frac{h^2}{\mu^2}\left(\dfrac{1}{6} - S(h\mu)\right), \\[0.5em]
    a_5 & = \frac{h^2}{\mu^3}\left(\frac{1}{24} - C(h\mu)\right).
\end{align*}
For $\varphi_k = \varphi_3$ the polynomial coefficients are:
\begin{align*}
    a_0 & = \frac{1}{6}, \\[0.5em]
    a_1 & = \frac{h}{24}, \\[0.5em]
    a_2 & = \frac{h^2}{120}, \\[0.5em]
    a_3 & = \frac{h^3}{720}, \\[0.5em]
    a_4 & = \frac{1}{\mu^4}\left(S(h\mu) - \dfrac{1}{6} + \dfrac{(h\mu)^2}{120}\right), \\[0.5em]
    a_5 & = \frac{1}{\mu^5}\left(C(h\mu) - \dfrac{h\mu}{24} + \frac{(h\mu)^{3}}{720}\right).
\end{align*}

If $R \neq 0$ and $T = 0$, then the characteristic polynomial reduces to
\[
z^6 + z^4\,P + z^2\,R = z^2(z^4 + z^2\,P + R).
\]
In this case, the eigenvalues are
\[
z = 0, 0, \pm i\,\mu, \pm i\,\nu,
\]
where
\[
\mu = \sqrt{\frac{P + \sqrt{P^2 - 4R}}{2}} \quad\text{and}\quad \nu = \sqrt{\frac{P - \sqrt{P^2 - 4R}}{2}}.
\]
For $\varphi_k = \varphi_1$ the polynomial coefficients are:
\begin{align*}
    a_0 & = 1, \\
    a_1 & = \frac{h}{2}, \\[0.5em]
    a_2 & = \frac{h^2}{\mu^2 - \nu^2}\left(\mu^2 S(h\nu) - \nu^2 S(h\mu)\right), \\[0.5em]
    a_3 & = \frac{h^2}{\mu^2 - \nu^2}\left(\dfrac{\mu^2}{\nu} C(h\nu) - \dfrac{\nu^2}{\mu} C(h\mu)\right), \\[0.5em]
    a_4 & = \frac{h^2}{\mu^2 - \nu^2}\left(S(h\nu) - S(h\mu)\right), \\[0.5em]
    a_5 & = \frac{h^2}{\mu^2 - \nu^2}\left(\dfrac{1}{\nu}C(h\nu) - \dfrac{1}{\mu}C(h\mu)\right).
\end{align*}
For $\varphi_k = \varphi_3$ the polynomial coefficients are:
\begin{align*}
    a_0 & = \frac{1}{6}, \\[0.5em]
    a_1 & = \frac{h}{24}, \\[0.5em]
    a_2 & = \frac{1}{\mu^2 - \nu^2}\left(\frac{\mu^2}{\nu^2}\left(\frac{1}{6} - S(h\nu)\right) - \frac{\nu^2}{\mu^2}\left(\frac{1}{6} - S(h\mu)\right)\right), \\[0.5em]
    a_3 & = \frac{1}{\mu^2 - \nu^2}\left(\frac{\mu^2}{\nu^3}\left(\frac{h\nu}{24} - C(h\nu)\right) - \frac{\nu^2}{\mu^3}\left(\frac{h\mu}{24} - C(h\mu)\right)\right), \\[0.5em]
    a_4 & = \frac{1}{\mu^2 - \nu^2}\left(\frac{1}{\nu^2}\left(\frac{1}{6} - S(h\nu)\right) - \frac{1}{\mu^2}\left(\frac{1}{6} - S(h\mu)\right)\right), \\[0.5em]
    a_5 & = \frac{1}{\mu^2 - \nu^2}\left(\frac{1}{\nu^3}\left(\frac{h\nu}{24} - C(h\nu)\right) - \frac{1}{\mu^3}\left(\frac{h\mu}{24} - C(h\mu)\right)\right).
\end{align*}

If $R, T \neq 0$, then the characteristic polynomial can be expressed as a cubic form
\[
w^3 + w^2\,P + w\,R + T, \qquad\text{where } w = z^2.
\]
The real root is given by
\[
w_1 = \rho\left(\frac{1}{3}P^2 - R\right) + \frac{1}{3}\left(\frac{1}{\rho} - P\right),
\]
where
\[
\rho = \left(\frac{2}{9PR - 2P^3 - 27T + 3\sqrt{3(4R^3 - P^2R^2 + 4P^3T - 18PRT + 27T^2}}\right)^{1/3}.
\]
The remaining two roots are given by
\[
w_2 = -\frac{1}{2}(b + \delta) \qquad\text{and}\qquad w_3 = \frac{1}{2}(\delta - b),
\]
where
\[
b = P + \lambda^2 \qquad\text{and}\qquad \delta = \sqrt{b - 4(R + \lambda^2 b)}.
\]

Define the following:
\[
\lambda = \sqrt{-w_1}, \quad \mu = \sqrt{-w_2}, \quad\text{and}\quad \nu = \sqrt{-w_3}.
\]
Then the interpolation polynomial coefficients for $\varphi_k = \varphi_1$ are:
\begin{align*}
    a_0 & = \frac{1}{\Delta_1\Delta_2\Delta_3}\left(\lambda^2\nu^2\Delta_3\frac{\sin(h\mu)}{h\mu} - \lambda^2\mu^2\Delta_2\frac{\sin(h\nu)}{h\nu} + \mu^2\nu^2\Delta_1\frac{\sin(h\lambda)}{h\lambda}\right), \\[0.5em]
    a_1 & = \frac{1}{\Delta_1\Delta_2\Delta_3}\left(\lambda^2\nu^2\Delta_3\frac{1 - \cos(h\mu)}{h\mu} - \lambda^2\mu^2\Delta_2\frac{1 - \cos(h\nu)}{h\nu} + \mu^2\nu^2\Delta_1\frac{1 - \cos(h\lambda)}{h\lambda}\right), \\[0.5em]
    a_2 & = \frac{1}{\Delta_1\Delta_2\Delta_3}\left((\nu^4 - \lambda^4)\Delta_3\frac{\sin(h\mu)}{h\mu} - (\mu^4 - \lambda^4)\Delta_2\frac{\sin(h\nu)}{h\nu} + (\mu^4 - \nu^4)\Delta_1\frac{\sin(h\lambda)}{h\lambda}\right), \\[0.5em]
    a_3 & = \frac{1}{\Delta_1\Delta_2\Delta_3}\bigg((\nu^4 - \lambda^4)\Delta_3\frac{1 - \cos(h\mu)}{h\mu} - (\mu^4 - \lambda^4)\Delta_2\frac{1 - \cos(h\nu)}{h\nu} \\
    & \phantom{= \frac{1}{\Delta_1\Delta_2\Delta_3} \bigg(} \quad + (\mu^4 - \nu^4)\Delta_1\frac{1 - \cos(h\lambda)}{h\lambda}\bigg), \\[0.5em]
    a_4 & = \frac{1}{\Delta_1\Delta_2\Delta_3}\left(\Delta_3\frac{\sin(h\mu)}{h\mu} - \Delta_2\frac{\sin(h\nu)}{h\nu} + \Delta_1\frac{\sin(h\lambda)}{h\lambda}\right), \\[0.5em]
    a_5 & = \frac{1}{\Delta_1\Delta_2\Delta_3}\left(\Delta_3\frac{1 - \cos(h\mu)}{h\mu} - \Delta_2\frac{1 - \cos(h\nu)}{h\nu} + \Delta_1\frac{1 - \cos(h\lambda)}{h\lambda}\right),
\end{align*}
where
\[
\Delta_1 = \lambda^2 - \mu^2, \quad \Delta_2 = \lambda^2 - \nu^2, \quad\text{and}\quad \Delta_3 = \mu^2 - \nu^2.
\]

The interpolation polynomial coefficients for $\varphi_k = \varphi_3$ are:
\begin{align*}
    a_0 & = \frac{1}{\Delta_1\Delta_2\Delta_3}\left(\lambda^2\nu^2\Delta_3 S(h\mu) - \lambda^2\mu^2\Delta_2 S(h\nu) + \mu^2\nu^2\Delta_1 S(h\lambda)\right), \\[0.5em]
    a_1 & = \frac{1}{\Delta_1\Delta_2\Delta_3}\left(\lambda^2\nu^2\Delta_3 \frac{C(h\mu)}{\mu} - \lambda^2\mu^2\Delta_2 \frac{C(h\nu)}{\nu} + \mu^2\nu^2\Delta_1 \frac{C(h\lambda)}{\lambda}\right), \\[0.5em]
    a_2 & = \frac{1}{\Delta_1\Delta_2\Delta_3}\left((\nu^4 - \lambda^4)\Delta_3 S(h\mu) - (\mu^4 - \lambda^4)\Delta_2 S(h\nu) + (\mu^4 - \nu^4)\Delta_1 S(h\lambda)\right), \\[0.5em]
    a_3 & = \frac{1}{\Delta_1\Delta_2\Delta_3}\left((\nu^4 - \lambda^4)\Delta_3 \frac{C(h\mu)}{\mu} - (\mu^4 - \lambda^4)\Delta_2 \frac{C(h\nu)}{\nu} + (\mu^4 - \nu^4)\Delta_1 \frac{C(h\lambda)}{\lambda}\right), \\[0.5em]
    a_4 & = \frac{1}{\Delta_1\Delta_2\Delta_3}\left(\Delta_3 S(h\mu) - \Delta_2 S(h\nu) + \Delta_1 S(h\lambda)\right), \\[0.5em]
    a_5 & = \frac{1}{\Delta_1\Delta_2\Delta_3}\left(\Delta_3 \frac{C(h\mu)}{\mu} - \Delta_2 \frac{C(h\nu)}{\nu} + \Delta_1 \frac{C(h\lambda)}{\lambda}\right).
\end{align*}

\end{document}